\documentclass{elsarticle}
\usepackage{graphics}
\usepackage{epsfig}
\usepackage{amsmath}
\usepackage{amssymb}
\usepackage{slashed}
\begin{document}

\begin{frontmatter}

\title{Holographic fermions with running chemical potential and dipole coupling}

\author[shu]{Li Qing Fang}
\ead{flqthunder@163.com}

\author[shu]{Xian-Hui Ge\corref{cor1}}
\ead{gexh@shu.edu.cn}

\author[sjtu]{Xiao-Mei Kuang}
\ead{xmeikuang@gmail.com}

\cortext[cor1]{Corresponding author}

\address[shu]{Department of Physics, Shanghai University, 200444 Shanghai,
China}
\address[sjtu]{Department of Physics, Shanghai Jiao Tong University, 200240 Shanghai, China}

\begin{abstract}
We explore the properties of the holographic fermions in extremal $R$-charged  black hole background with a running chemical potential, as well as the dipole coupling between fermions and the gauge field in the bulk. We find that although the running chemical potential effect the location of the Fermi surface, it does not change the type of fermions. We also study the onset of the Fermi gap and the gap effected by running chemical potential and the dipole coupling. The spectral function in the limit $\omega\rightarrow0$ and the existence of the Fermi liquid are also investigated. The running chemical potential and the dipole coupling altogether can make a non-Fermi liquid become the Landau-Fermi type.
\end{abstract}


\end{frontmatter}

\section{Introduction}
The gauge/gravity correspondence\cite{Maldacena,Gubser-1,E.Witten} is known as a powerful tool to provide key insights to disclose the mysterious phenomena observed in high temperature superconductor and heavy fermion systems, which are relative to the strongly correlated electron system. People are still lack of a general theoretical framework to deal with this so called non-Fermi liquids. Fortunately, with the use of gauge/gravity correspondence, tremendous progress has been made
in our understanding of these strongly coupled system. In \cite{f1,f2,IL,f3,f4,f5}, the authors has investigated the holographic Fermi surfaces where Fermi liquid and non-Fermi liquid are obtained by probing the behavior of the Dirac field in the RN-AdS black hole.
\begin{figure}
 \centering
 \includegraphics[width=.5\textwidth]{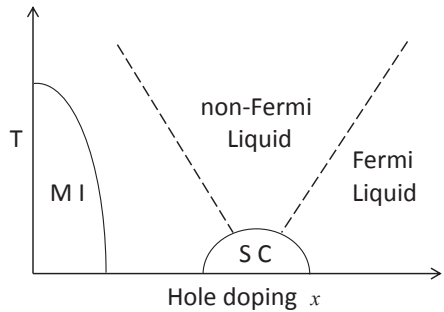}
 \caption{A schematic plot of the phase diagram of the cuprate superconductors with hole doping.}
 \label{phasediagram}
\end{figure}

The high $T_c$ cuprates have a very rich phase structure and at different doping region show their very different properties. As shown in figure \ref{phasediagram}, we can see that without doping ($x=0$) the cuprate is the antiferromagnetic Mott insulator.
By doping holes, the antiferromagnetic Mott phase is rapidly destroyed. Beyond some critical doping $x_c$, the  non-Fermi  liquid phase and superconducting ground state emerge. At the over-doped region, the transition between the Fermi-liquid phase and the non-Fermi liquid phase occurs gradually by crossover. The dopant $x$ actually changes the chemical potential of the system. As the chemical potential running, the structure of the Fermi surfaces
varies and new phases appear. So it is proper to interpret  the chemical potential $\mu$ in the gravity theory as the doping $x$ in the cuprates. It is interesting to investigate the holographic fermions with running chemical potential ( i.e. varying ``doping"). In this set up, we need  to  consider two chemical potentials, one as the background and the other as ``doping".

The problem now is how to include the ``doping" in the gauge/gravity set up.  So, it is proper to  consider R-charged black hole solution of the STU model of gauged supergravity  which  contains three different charges $Q_a$, $Q_b$ and $Q_c$\cite{Rcharge1,Rcharge2,GubserBackground}. For our purpose of discussing the ``doping"  effect, we consider a special condition of R-charged black hole by setting two charges equal, i.e., $Q_b=Q_c=Q_2$ and $Q_a=Q_1$. Such a two
charge black hole solution can be derived from the effective Lagrangian \cite{GubserBackground}
 \begin{eqnarray}\label{action}
e^{-1}\mathcal{L}=-R+\frac{1}{2}(\partial\phi)^{2}+\frac{8}{L^{2}}e^{\frac{\phi}{\sqrt{6}}}
&+&\frac{4}{L^{2}}e^{\frac{-2\phi}{\sqrt{6}}}
-e^{\frac{-4\phi}{\sqrt{6}}}f_{\mu\nu}f^{\mu\nu} \nonumber\\
&-&2e^{\frac{2\phi}{\sqrt{6}}}F_{\mu\nu}F^{\mu\nu}
-2\epsilon^{\mu\nu\rho\sigma\tau}f_{\mu\nu}F_{\rho\sigma}A_{\tau}
\end{eqnarray}
where $f_{\mu\nu}$ and $F_{\mu\nu}$ are the field strength of two gauge fields while $\phi$ is scalar field, which are associated with abelian gauge field $A_{\tau}$.  In order to study how the chemical potentials effect on the holographic fermions directly, we consider only one of the gauge fields couples with the bulk fermion and leave the other gauge field as the background. In the zero temperature limit, $Q_1$ and $Q_2$ satisfy a fixed relation and we can define a charge ratio. As the charge ratio  as well as the chemical potential ratio varies, the corresponding two chemical potentials change.

The main motivation of the present paper is to obtain the complete phase diagram of the holographic system with a varying chemical potential. We will check whether it makes a non-Fermi liquid become the Landau-Fermi type as the chemical potential runs. We will work in the probe limit, such that the backreaction of the Dirac field on the background geometry can be neglected.
Note that there are two main drawbacks by using the probe approximation to realize the property of holographic fermions. Firstly, it might be difficult to calculate some quantities from large N limit. Secondly, it is hard to have a precise description of the field aspect and interactions of the dual field theory. But in order to well mimic to the condensed matter physics, a phenomenological approach is still interesting. For example, the holographic investigation with the use of the probe approximation in R-charge black hole background was studied in \cite{R1,R2}. In what follows, we will summarize our main results as follows:
\begin{itemize}
  \item
 We will first explore the holographic fermion system in the two charges R-charged black hole background without dipole coupling. Both for  the cases that only $f_{\mu\nu}$ or $F_{\mu\nu}$  couples with fermions, varying chemical potential changes the position of the Fermi surface and the dispersion relation, but it can not affect the type of the dual liquid. This implies that running chemical potential
  might not be able to play the same role of ``doping"  as in real cuprates \cite{wen}.
  \item
  We will then  consider both  the dipole coupling  and running chemical potential effects on the holographic fermions. For only $f_{\mu\nu}$   couples with fermions, we find that the bigger chemical potential $\mu_1$ leads to the generation of the dynamical gap easier. But for only $F_{\mu\nu}$   couples with fermions case, the critical dipole coupling constant $p_{2cri}$ is a quadratic relation with the chemical potential $\mu_2$. The oscillating regions are also observed. We find that the running chemical potential can indeed influence on the type of dual liquid in the presence of  the dipole coupling. The phase structure is greatly enriched with both the running chemical potential and the dipole coupling.
\end{itemize}

Now, we would like to briefly review some relevant literature on holographic fermions and clarify the differences between our results with these literature. Probing holographic fermions in single charged RN-AdS black hole were studied in \cite{f2,f3} with a fixed chemical potential at a fixed temperature. But we will work in the multi-charged case in which chemical potentials can change at zero temperature limit. The observation of the transferred spectral density and the emerged Mott gap  were archived by introducing the coupling between the fermion and gauge field through a dipole interaction in the bulk \cite{R.G.Leigh1}. There exists a critical coupling strength at which a Mott gap opens up and continuously increasing the coupling strength makes the gap wider. More details of how the Fermi systems effect by varying the dipole coupling was also explored in \cite{R.G.Leigh2,dipole3}. Note that dipole coupling term  can also produce the similar phase diagram with that of the cuprates, but the physical meaning of dipole coupling strength needs future investigations.  The  bulk dipole coupling effect was also discussed in the top-down   string-embedded  model \cite{GSW1,GSW2}. The extended investigation on the dipole coupling in other gravity theory such as dilatonic gravity and Gauss-Bonnet gravity also can be seen in\cite{JPWu1,Wen,Kuang1}. On the other hand, the implementation of the holographic non-relativistic fermionic fixed points by imposing Lorentz breaking boundary conditions was proposed in \cite{D.Tong1}. This makes an infinite flat band appear in the boundary field theory. The related studies on the holographic non-relativistic fermionic fixed points have been reported in \cite{D.Tong2,WJL1,WJL2,Gursoy1,AMM,Fang,JPWu2,Gursoy2,Kuang2}. In the holographic set up, the dipole coupling term  seems to  act as the role of ``doping" in the high temperature cuprate in terms of the Hubbard model.  But the physical origin of the dipole interaction term calls for further investigation. Moreover, the running chemical potential was not considered in the motioned references.  In \cite{GubserBackground}, a special R-charged black holes background with two distinct gauge fields were constructed  and  the properties of the holographic fermions were discussed from the top-down point of view.  But the authors did not consider the situation that  only one gauge field ($f_{\mu\nu}$ or $F_{\mu\nu}$) couples with the fermions.

The organization of this paper is as follows. In section 2, we will give a brief review  on the a special R-charged black hole solution which we will use in following discussion. In section 3, we will present the holographic setup of the dual fermion  system in the probe setup. Then, we will study the influences on the Fermi surface, the type of Fermi liquid and the Fermi
gaps, due to running chemical potentials and the dipole coupling in section 4.
Conclusions and discussions will be presented in section 5.
\section{R-charged  Black Hole Background}
The background we will use in our paper,  is the extremal  R-charged  black holes  obtained in \cite{GubserBackground} with two distinct chemical potentials. For our purpose of studying running chemical potential effect, we will consider the behavior of fermions in this background and later with the dipole coupling.

\subsection{Review on the R-charged  black hole}
In this section, we will give a brief review on the R-charged  black hole solution in $4+1$-dimensional spacetime. The R-charged black hole \cite{Rcharge1,Rcharge2} is a solution of STU model of gauged supergravity. The ten dimensional type IIB string theory background on $AdS_{5}\times S^{5}$ provides gravity dual of a $\mathcal{N}=4$ super Yang-Mills theory. The dimensional reduction on the $S^{5}$ which leads to an SO(6) isometry group is correspond to charged black hole with three U(1) charges. This isometry group represents the
R-symmetry group in the gauge theory. In our paper, we will only use a special case, which contain two equal charge and one unequal charge. The metric is given by \cite{GubserBackground}
\begin{eqnarray}\label{metric}
~~~~~~~~~
ds^2&=&-g_{tt}dt^2+g_{xx} d x^2+g_{yy} d y^2+g_{zz} d z^2+g_{rr}dr^2 \nonumber\\
&=&e^{2A(r)}[h(r)dt^2-dx^2-dy^2-dz^2]-\frac{e^{2B(r)}}{h(r)}dr^2,\nonumber\\
A_{\mu}dx^{\mu}&=&\Phi_{1}(r)dt,~~~~~B_{\mu}dx^{\mu}=\Phi_{2}(r)dt,~~~~~\phi=\phi(r)
\end{eqnarray}
with
\begin{eqnarray}\label{BHsolution}
~~~~~~~~~~~~
A(r)&=&\log(\frac{r}{L})+\frac{1}{6}\Big(1+\frac{Q_{1}^{2}}{r^{2}}\Big)
+\frac{1}{3}\Big(1+\frac{Q_{2}^{2}}{r^{2}}\Big),\nonumber\\
B(r)&=&-\log(\frac{r}{L})-\frac{1}{3}\Big(1+\frac{Q_{1}^{2}}{r^{2}}\Big)
-\frac{2}{3}\Big(1+\frac{Q_{2}^{2}}{r^{2}}\Big),\nonumber\\
h(r)&=&1-\frac{r^{2}(r_{H}^{2}+Q_{1}^{2})(r_{H}^{2}+Q_{2}^{2})^{2}}
{r_{H}^{2}(r^{2}+Q_{1}^{2})(r^{2}+Q_{2}^{2})^{2}},\nonumber\\
\phi(r)&=&-\sqrt{\frac{2}{3}}\log\Big(1+\frac{Q_{1}^{2}}{r^{2}}\Big)
+\sqrt{\frac{2}{3}}\log\Big(1+\frac{Q_{2}^{2}}{r^{2}}\Big),\nonumber\\
\Phi_{1}(r)&=&\frac{Q_{1}(r_{H}^{2}+Q_{2}^{2})}{2L r_{H}\sqrt{r_{H}^{2}+Q_{1}^{2}}}\Big(1-\frac{r_{H}^{2}+Q_{1}^{2}}{r^{2}+Q_{1}^{2}}\Big),\nonumber\\
\Phi_{2}(r)&=&\frac{Q_{2}\sqrt{r_{H}^{2}+Q_{1}^{2}}}{2L r_{H}}\Big(1-\frac{r_{H}^{2}+Q_{2}^{2}}{r^{2}+Q_{2}^{2}}\Big).
\end{eqnarray}
As discussed in \cite{GubserBackground}, when $Q_1=Q_2$, the solution is a ``3-charge" black hole and it is a ``1-charge" black hole\footnote{``3-charge" and ``1-charge" black hole are the special situation of R-charged black hole, which are the terminology used in \cite{GubserBackground}.} when $Q_2=0$.
In this work, we will focus on the R-charged black hole with general $Q_1$ and $Q_2$ in (\ref{BHsolution}) where the horizon $r_H$ satisfies
$h(r_H)=0$. Besides, the black hole background is the anti-de Sitter space in the large $r$ limit.

The Hawking temperature of the black hole is
\begin{equation}
T=\frac{2r_{H}^{4}+Q_{1}^{2}r_{H}^{2}-Q_{1}^{2}Q_{2}^{2}}{2\pi L^{2}r_{H}^{2}\sqrt{r_{H}^{2}+Q_{1}^{2}}},
\end{equation}
which will also be considered as the temperature of the system in the boundary.
The chemical potentials relative to the two gauge fields read as
\begin{equation}
\mu_{1}=\frac{Q_{1}(r_{H}^{2}+Q_{2}^{2})}{L^{2} r_{H}\sqrt{r_{H}^{2}+Q_{1}^{2}}},~~~\mu_{2}
=\frac{\sqrt{2}Q_{2}\sqrt{r_{H}^{2}+Q_{1}^{2}}}{L^{2}r_{H}},
\end{equation}
respectively. Moreover, the corresponding charge densities are
\begin{equation}
\rho_{1}=\frac{Q_{1}s}{2\pi r_{H}}~~~{\rm and}~~~\rho_{2}=\frac{\sqrt{2}Q_{2}s}{2\pi r_{H}}
\end{equation}
where $s$ is the entropy density with the form
\begin{equation}
s=\frac{1}{4GL^{3}}(r_{H}^{2}+Q_{1}^{2})^{1/2}(r_{H}^{2}+Q_{2}^{2}).
\end{equation}
The entropy density obtained is nonzero at the zero temperature.
We are interesting in the extremal black hole case, i.e., the solution with zero temperature but nonzero chemical potential. Beginning from the extremal condition, we can express $Q_2$ with $Q_1$ and $r_H$ as
\begin{equation}\label{extremal}
Q_{2}^{2}=\frac{2r_{H}^{4}}{Q_{1}^{2}}+r_{H}^{2}.
\end{equation}
To proceed, we define a dimensionless charge ratio for convenience
\begin{equation}\label{QR}
Q_{R} \equiv\frac{Q_{1}}{Q_{2}}.
\end{equation}
By combining (\ref{extremal}) and (\ref{QR}), we can re-express the two charges in terms of $Q_R$ as
\begin{eqnarray}
~~~~~~
Q_{1}&=&\frac{r_{H}}{\sqrt{2}}\sqrt{(8Q_{R}^{2}+Q_{R}^{4})^{1/2}+Q_{R}^{2}},\nonumber\\
Q_{2}&=&\frac{r_{H}}{\sqrt{2}Q_{R}}\sqrt{(8Q_{R}^{2}+Q_{R}^{4})^{1/2}+Q_{R}^{2}}.
\end{eqnarray}

\begin{figure}
\centering
\includegraphics[width=.46\textwidth]{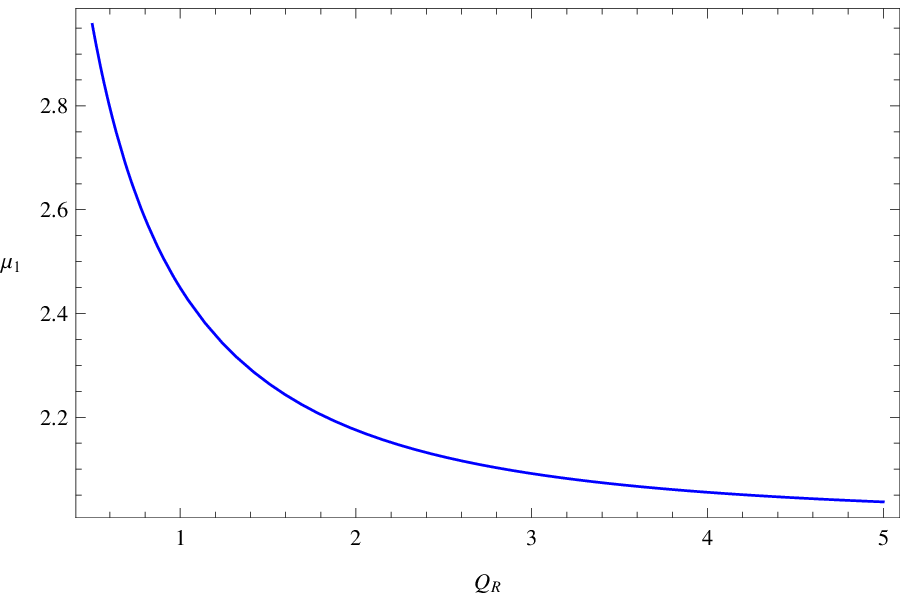}\hspace{0.6cm}
\includegraphics[width=.46\textwidth]{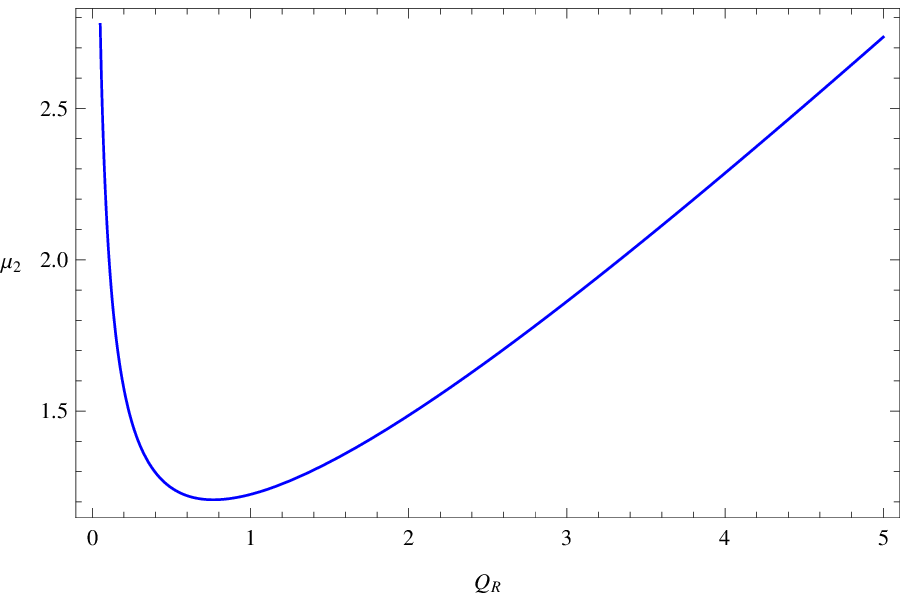}
\caption{The plots of $\mu_1$ (left) and $\mu_2$ (right) versus $Q_R$.}
\label{muQR}
\end{figure}

In the following, we will set the horizon radius $r_H=1$ and investigate the system with running chemical potentials\footnote{Note that the ratio of the chemical potentials, which characterizes the unbalance between the ¡°background¡± chemical potential and the ``doping" , depends on  $Q_R$. The explicit relation between the two chemical potentials at zero temperature as a function of $Q_R$ is given by $\frac{\mu_1}{\mu_2}=\sqrt{Q_R^2+8}-Q_R$. } $\mu_1$ and $\mu_2$ by varying the value  of $Q_R$. The left plot of figure \ref{muQR} shows  us that $\mu_1$ monotonously decreases as $Q_R$ increases, while $\mu_2$ is a quadratic function of $Q_R$ and reaches its minimum at $Q_{R}=\sqrt{2-\sqrt{2}}\sim0.765$ in the right plot of figure \ref{muQR}. So we see that changing $Q_R$ corresponds to varying chemical potentials.  We also note that $Q_R=1$ and $Q_R=\infty$ correspond to the ``3-charge" black hole and the ``1-charge" black hole, respectively.

\subsection{Near-horizon limit}
We move on to analyze the near horizon limit of the extremal R-charged black hole background. Similar to the geometry discussed in \cite{f3}, near the horizon region, the metric present a double pole and the geometry approaches $AdS_{2}\times \mathbb{R}^{3}$ in the zero temperature limit. Specially, in the extremal
case, the metric  near the horizon has its behaviors \cite{GubserBackground}
\begin{eqnarray}
~~~~~~
&g_{tt}\rightarrow -\tau_{0}^{2}(r-r_{H})^{2},~~~g_{rr}\rightarrow \frac{L_{2}^{2}}{(r-r_{H})^{2}},~~~g_{xx}, g_{yy}, g_{zz}\rightarrow k_{0}^{2},\nonumber\\
&\Phi_{i}\rightarrow\beta_{i}(r-r_{H}),~~~\phi\rightarrow\phi_{0},
\end{eqnarray}
where
\begin{eqnarray}\label{ads2coefficient}
~~~~~~~~
\tau_{0}=\frac{2^{1/3}}{L}\Big(\frac{Q_{1}}{r_{H}}\Big)^{1/3}
\sqrt{\frac{4r_{H}^{2}+Q_{1}^{2}}{r_{H}^{2}+Q_{1}^{2}}},~~~~
L_{2}=\frac{L}{2^{2/3}}\frac{Q_{1}^{1/3}r_{H}^{2/3}}{\sqrt{4r_{H}^{2}+Q_{1}^{2}}},\nonumber\\
k_{0}=\frac{2^{1/3}}{L}\Big(\frac{r_{H}}{Q_{1}}\Big)^{2/3}\sqrt{r_{H}^{2}+Q_{1}^{2}},~~~~~~
\phi_{0}=\sqrt{\frac{2}{3}}\log\Big(\frac{2r_{H}^{2}}{Q_{1}^{2}}\Big),~~\nonumber\\
\beta_{1}=\frac{2r_{H}^{2}}{L Q_{1}\sqrt{r_{H}^{2}+Q_{1}^{2}}},~~~~~~~~~~~~~~~~
\beta_{2}=\frac{Q_{1}\sqrt{2r_{H}^{2}+Q_{1}^{2}}}{2L r_{H}\sqrt{r_{H}^{2}+Q_{1}^{2}}}.~
\end{eqnarray}
Then, we make the new transformations
\begin{equation}\label{scaling}
r-r_{H}=\frac{\lambda L_{2}^{2}}{\tau_{0}\zeta},~~~t=\lambda^{-1}\tau.
\end{equation}
Considering the scaling limit $\lambda\rightarrow 0$ with finite fixed $\zeta$ and $\tau$,
we find that the metric have the formulation of $AdS_{2}\times \mathbb{R}^{3}$
\begin{equation}\label{extremalads2}
ds^2=\frac{L_{2}^{2}}{\zeta^{2}}(-d\tau^2+d\zeta^2)+k_{0}^{2}(dx^2+dy^2+dz^2)
\end{equation}
with $L_2$ the $AdS_2$ length. Meanwhile, the two gauge fields are re-scaled into the form
\begin{equation}
A_{\mu}dx^{\mu}=\frac{e_{1}}{\zeta}d\tau,~~~B_{\mu}dx^{\mu}=\frac{e_{2}}{\zeta}d\tau.
\end{equation}
The coefficients $e_i(i=1,2)$ of the gauge field in the $AdS_2$ limit has the expression $e_i=\frac{\beta_{i}L_{2}}{\tau_{0}}$ with $\beta{i}$ and
$\tau_0$ defined in (\ref{ads2coefficient}). The above near horizon behavior of the background solution plays very important role when we discuss the low energy
behavior of the holographic system, which will be shown in what follows.

\section{Holographic Setup}
In this section, we will set up the probe Fermi model in the above background.  We first write down  the Dirac equation in the bulk theory with the  dipole coupling term. Then,
we discuss the low frequency behavior of the Fermi system by using the analytical method.

\subsection{Equations of motion in the bulk theory}
In order to study the running chemical potential effect with dipole coupling  of fermions in the dual boundary theory, we consider a probe Dirac fermion taking the mass $m$ and charge $q$ with the $4+1$-dimensional bulk action
\begin{equation}\label{bulkfermionaction}
S_{bulk}=\int
d^{4+1}x\sqrt{-g}i\bar{\psi}\Big(\Gamma^a D_{a}-
m-i\delta_{1i}p_i\slashed{f}-i\delta_{2i}p_i\slashed{F}\Big)\psi,
\end{equation}
where  $\Gamma^a=(e_\mu)^a\Gamma^\mu$, $\slashed{f}=\frac{1}{4}\Gamma^{\mu\nu}(e_\mu)^{a}(e_\nu)^{b}f_{ab}$, $\slashed{F}=\frac{1}{4}\Gamma^{\mu\nu}(e_\mu)^{a}(e_\nu)^{b}F_{ab}$,
 \textbf{ $\delta_{ji} (i,j=1,2)$  } is the delta function  and  $p_i$ denotes the magnetic dipole couplings strength with the two gauge fields $A_{\mu}$ and $B_{\mu}$, respectively. The expression for the covariant derivative $D_{a}$ is
\begin{equation}\label{CovariantDerivative}
D_{a}=\partial_{a}+\frac{1}{4}(\omega_{\mu\nu})_a\Gamma^{\mu\nu}-i\delta_{1i}q_{i}A_{a}
-i\delta_{2i}q_{i}B_{a},
\end{equation}
with  $\Gamma^{\mu\nu}=\frac{1}{2}[\Gamma^\mu,\Gamma^\nu]$ and  the spin connection $(\omega_{\mu\nu})_{a}=(e_\mu)^b\nabla_a(e_\nu)_b$, where
$(e_\mu)^{a}$ forms a set of orthogonal normal vector bases \cite{Conventions1}.

The Dirac equation derived from the above action reads
\begin{equation}\label{Diraceom1}
\Gamma^a D_{a}\psi-m\psi-i\delta_{1i}p_i\slashed{f}\psi-i\delta_{2i}p_i\slashed{F}\psi=0.
\end{equation}
In order to remove the spin connection in the Dirac equation and to investigate in Fourier space,
following \cite{f2}, we make a transformation $\psi=(-g g^{rr})^{-\frac{1}{4}}e^{-i\omega t+ik_{i}x^{i}}\phi$.
Considering the rotation symmetry in the spatial directions, we can simply set  $k_x=k$. The Dirac
equation (\ref{Diraceom1}) has the form
\begin{eqnarray}\label{phieom}
~~~
\big[\frac{\sqrt{g_{xx}}}{\sqrt{g_{rr}}}\Gamma^r\partial_r-\frac{\sqrt{g_{xx}}}{\sqrt{g_{tt}}}\Gamma^{t}
i(\omega+\delta_{1i}q_{i}A_t+\delta_{2i}q_{i}B_t)-\sqrt{g_{xx}}m+ik\Gamma^{x}\nonumber\\
-\frac{i\delta_{1i}p_{i}}{2}\frac{\sqrt{g_{xx}}}{\sqrt{g_{tt}g_{rr}}}\Gamma^{rt}\partial_{r}A_{t}
-\frac{i\delta_{2i}p_{i}}{2}\frac{\sqrt{g_{xx}}}{\sqrt{g_{tt}g_{rr}}}\Gamma^{rt}\partial_{r}B_{t}
\big]\phi=0.
\end{eqnarray}
It is obvious that the equation above only depends on three gamma matrices $\Gamma^{r}, \Gamma^{t}, \Gamma^{x}$. So it is convenient
to set $\phi=\left(\begin{array}{c}\phi_1 \\ \phi_2\\ \end{array} \right)$ and consider the following basis for gamma matrices \cite{Photoemission}:
\begin{eqnarray}
\label{GammaMatrices}
 && \Gamma^{r} = \left( \begin{array}{cc}
-\sigma^3   & 0  \\
0 & -\sigma^3
\end{array} \right), \;\;
 \Gamma^{t} = \left( \begin{array}{cc}
 i \sigma^1   & 0  \\
0 & i \sigma^1
\end{array} \right),  \;\;
\Gamma^{x} = \left( \begin{array}{cc}
-\sigma^2   & 0  \\
0 & \sigma^2
\end{array} \right),
 \;\;
\cdots
\end{eqnarray}
By taking the above operations, we derive the equation of motion (\ref{phieom})
\begin{eqnarray}\label{phi12eom}
&&\frac{\sqrt{g_{xx}}}{\sqrt{g_{rr}}}\partial_r
\left(\begin{array}{c}\phi_1 \\ \phi_2\\ \end{array} \right)
+\sqrt{g_{xx}}m\sigma^{3}\otimes
\left(\begin{array}{c}\phi_1 \\ \phi_2\\ \end{array} \right)\nonumber\\
&&=\frac{\sqrt{g_{xx}}}{\sqrt{g_{tt}}}(\omega+\delta_{1i}q_{i} A_{t}+\delta_{2i}q_{i} B_{t})i\sigma^{2}\otimes
\left(\begin{array}{c}\phi_1 \\ \phi_2\\ \end{array} \right)
\mp k\sigma^{1}\otimes
\left(\begin{array}{c}\phi_1 \\ \phi_2\\ \end{array} \right)\nonumber\\
&&-\frac{\delta_{1i}p_i\sqrt{g_{xx}}}{\sqrt{g_{tt}g_{rr}}}\partial_{r}A_{t}\sigma^{1}\otimes
\left(\begin{array}{c}\phi_1 \\ \phi_2\\ \end{array} \right)
-\frac{\delta_{2i}p_i\sqrt{g_{xx}}}{\sqrt{g_{tt}g_{rr}}}\partial_{r}B_{t}\sigma^{1}\otimes
\left(\begin{array}{c}\phi_1 \\ \phi_2\\ \end{array} \right).
\end{eqnarray}

Furthermore, by setting $\phi_I=\left(\begin{array}{c}y_I \\ z_I\\\end{array} \right)$ and defining $\xi_I=\frac{y_I}{z_I}$,
we decouple the equation of motion and thus reduce to
\begin{equation}\label{floweqa}
(\frac{\sqrt{g_{xx}}}{\sqrt{g_{rr}}}\partial_r
+2\sqrt{g_{xx}}m)\xi_{I}
=[v_{-}+(-1)^{I}k]+[v_{+}-(-1)^{I}k]\xi_{I}^{2},
\end{equation}
with  $v_{\pm}$ expressed as
\begin{eqnarray}
v_{\pm}=\frac{\sqrt{g_{xx}}}{\sqrt{g_{tt}}}(\omega+\delta_{1i}q_{i} A_{t}+\delta_{2i}q_{i} B_{t})
\pm\frac{\delta_{1i}p_i\sqrt{g_{xx}}}{\sqrt{g_{tt}g_{rr}}}\partial_{r}A_{t}
\pm\frac{\delta_{2i}p_i\sqrt{g_{xx}}}{\sqrt{g_{tt}g_{rr}}}\partial_{r}B_{t}.
\end{eqnarray}

To solve the Dirac equation, we need impose the boundary condition. Near the AdS boundary $r\rightarrow\infty$, from the above equation, we can deduce that $\phi_{I}$ behave as
\begin{equation}\label{bdysol}
\phi_I\rightarrow a_I r^{-m}\left(
\begin{array}{c}
1 \\0 \\ \end{array}
 \right)+b_Ir^{m}\left(
\begin{array}{c} 0 \\ 1 \\\end{array}\right).
\end{equation}
As discussed in \cite{IL}, if $a_{I}\left( \begin{array}{c}
1 \\0 \\ \end{array}\right)$
and $b_{I}\left(\begin{array}{c} 0 \\ 1 \\\end{array}\right)$ are related by
$a_{I}\left( \begin{array}{c} 1 \\ 0 \\\end{array}\right)
=\mathcal{S}b_{I}\left( \begin{array}{c} 0 \\ 1 \\\end{array}\right)$,
the boundary Green's functions $G(\omega,k)$ is given by $G=-i \mathcal{S}\gamma^{0}$.
Then the Green function can be expressed as
\begin{equation}\label{gi}
G(\omega,k)=\lim_{r\rightarrow\infty}r^{2m}
\left(
\begin{array}{cc}
\xi_1 & 0 \\
0 & \xi_2 \\
\end{array}
\right).
\end{equation}
Up to normalization, the fermion spectral function can be written as
\begin{equation}\label{definitionAwk}
A(\omega,k)\equiv \rm{Tr} [ImG(\omega,k)].
\end{equation}
On the other hand, at the horizon, we choose the in-falling boundary condition
\begin{equation}\label{bdycond}
\xi_I\mid_{r=r_{H}}= \left\{ \begin{array}{ll}
i & \textrm{~~for $\omega\neq0$}\\
\frac{mL_{2}-\nu_{k}^{I}}{\Big(\frac{\delta_{1i}q_{i}\beta_{i}L_{2}}{\tau_{0}}
+\frac{\delta_{2i}q_{i}\beta_{i}L_{2}}{\tau_{0}}\Big)+\tilde{m}_{I}L_{2}} & \textrm{~~for $\omega=0$.}
\end{array} \right.
\end{equation}
Here $\nu_{k}^{I}+\frac{1}{2}$ and $\tilde{m}_{I}$ is conformal dimension and effective mass in the low frequency theory dual to the $AdS_2$ background, which will be discussed in the following subsection.

\subsection{Low frequency behavior}
In this subsection, we will analyze the holographic system in the low frequency limit. For the extremal  black holes, the geometry approaches $AdS_{2}\times \mathbb{R}^{3}$ near horizon region controlled by (\ref{extremalads2}). In this region, we can express $\phi_{I}$ in terms of $\zeta$ and expand in powers of $\omega$ as\footnote{The low frequency limit $\omega\rightarrow0$ is equivalent to the the scaling limit $\lambda\rightarrow 0$ in (\ref{scaling}). We use $\omega$ instead of $\lambda$ for unity in this section. }
\begin{equation}\label{power}
\phi_{I}(\zeta)=\phi_{I}^{(0)}(\zeta)+\omega\phi_{I}^{(1)}(\zeta)+\omega^{2}\phi_{I}^{(2)}(\zeta)+\cdots
\end{equation}
By substituting (\ref{power}) into (\ref{phi12eom}), we obtain the leading order term
\begin{eqnarray}
\partial_{\zeta}\phi_{I}^{(0)}(\zeta)=\frac{L_{2}}{\zeta} m \sigma^{3}\phi_{I}^{(0)}(\zeta)
-i\Big(1+\frac{\delta_{1i}q_{i}\beta_{i}L_{2}}{\tau_{0}\zeta}
+\frac{\delta_{2i}q_{i}\beta_{i}L_{2}}{\tau_{0}\zeta}\Big)\sigma^{2}\phi_{I}^{(0)}(\zeta)\nonumber\\
+\frac{L_{2}}{\zeta}\Big[\frac{\delta_{1i}p_{i}\beta_{i}}{\tau_{0}L_{2}}
+\frac{\delta_{2i}p_{i}\beta_{i}}{\tau_{0}L_{2}}-(-1)^{I}\frac{k}{k_{0}}\Big]\sigma^{1}\phi_{I}^{(0)}(\zeta).
\end{eqnarray}
The above equation is identical to the equation  of motion in $AdS_2$
background for massive spinor
fields with masses $(m,\tilde{m}_{I})$ where
\begin{equation}
\tilde{m}_{I}=\frac{\delta_{1i}p_{i}\beta_{i}}{\tau_{0}L_{2}}
+\frac{\delta_{2i}p_{i}\beta_{i}}{\tau_{0}L_{2}}-(-1)^{I}\frac{k}{k_{0}}
\end{equation}
are the time-reversal violating mass terms. According
to the analysis in \cite{f3}, $\phi_{I}^{(0)}(\zeta)$ is dual to the spinor operators $\mathbb{O}_{I}$ in the IR $CFT_{1}$ with the
conformal dimensions  $\delta_{k}^{I}=\nu_{k}^{I}+\frac{1}{2}$, where
\begin{equation}\label{nuk}
\delta_{k}^{I}=\nu_{k}^{I}+\frac{1}{2}=\sqrt{m^{2}L_{2}^{2}+\tilde{m}_{I}^{2}L_{2}^{2}-\Big(\frac{\delta_{1i}q_{i}\beta_{i}L_{2}}{\tau_{0}}
+\frac{\delta_{2i}q_{i}\beta_{i}L_{2}}{\tau_{0}}\Big)^{2}}+\frac{1}{2}.
\end{equation}
As in \cite{f3}, the fermion retarded correlator at low frequency can be expressed with the retarded Green functions
of $\mathbb{O}_{I}$. By matching the inner $AdS_2$ and outer $AdS_5$ solutions, we can extract the coefficients $a_{I}$ and $b_{I}$ in (\ref{bdysol})
\begin{eqnarray}\label{match}
~~~~~~~~~~
a_{I}&=&(a_{I}^{(0)}+\omega a_{I}^{(1)}+\cdots)+(\tilde{a}_{I}^{(0)}
+\omega \tilde{a}_{I}^{(1)}+\cdots)\mathcal{G}_{k}^{I}(\omega),\nonumber\\
b_{I}&=&(b_{I}^{(0)}+\omega b_{I}^{(1)}+\cdots)+(\tilde{b}_{I}^{(0)}
+\omega \tilde{b}_{I}^{(1)}+\cdots)\mathcal{G}_{k}^{I}(\omega),
\end{eqnarray}
with $a_{I}^{(n)}, \tilde{a}_{I}^{(n)}, b_{I}^{(n)}$ and $\tilde{b}_{I}^{(n)}$ calculated numerically and $\mathcal{G}_{\alpha}(k,\omega)$
the retarded Green functions of the dual operators $\mathbb{O}_{I}$ given by
\begin{eqnarray}
~~~~~~
\mathcal{G}_{k}^{I}(\omega)=\Big[e^{-i\pi\nu_{k}^{I}}\frac{\Gamma(-2\nu_{k}^{I})\Gamma(1+\nu_{k}^{I}
-i\frac{\delta_{1i}q_{i}\beta_{i}L_{2}}{\tau_{0}}-i\frac{\delta_{2i}q_{i}\beta_{i}L_{2}}{\tau_{0}})}
{\Gamma(2\nu_{k}^{I})\Gamma(1-\nu_{k}^{I}-i\frac{\delta_{1i}q_{i}\beta_{i}L_{2}}{\tau_{0}}
-i\frac{\delta_{2i}q_{i}\beta_{i}L_{2}}{\tau_{0}})}\nonumber\\
\times\frac{(m+i\tilde{m}_{I})L_{2}-i\frac{\delta_{1i}q_{i}\beta_{i}L_{2}}{\tau_{0}}
-i\frac{\delta_{2i}q_{i}\beta_{i}L_{2}}{\tau_{0}}-\nu_{k}^{I}}
{(m+i\tilde{m}_{I})L_{2}-i\frac{\delta_{1i}q_{i}\beta_{i}L_{2}}{\tau_{0}}
-i\frac{\delta_{2i}q_{i}\beta_{i}L_{2}}{\tau_{0}}+\nu_{k}^{I}}\Big]\omega^{2\nu_{k}^{I}}.
\end{eqnarray}
It is worthwhile to emphasize that when $2\nu_{k}^{I}$ is an integer, (\ref{match}) is invalid and an additional terms should be considered\cite{f3}.

\section{Properties of the Fermion System with Running Chemical Potential and the Dipole Coupling}
We are going to explore the properties of the holographic system by numerically integrating the flow equation (\ref{floweqa}). We will calculate the fermion spectral function $A(\omega, k)$ defined in (\ref{definitionAwk}) and the density of states $A(\omega)$ by doing the integration of $A(\omega, k)$ over $k$.
Note that from figure \ref{muQR}, it is obvious that $\mu_1$ and $\mu_2$ depend  on $Q_R$ with different rules, so we will discuss both the fermi surface and the gap for the two cases that $A_t$ and $B_t$ couple to probe fermions separately. For the convenience of our numerical calculation, we will set $m=0$, $L=1$, $r_H=1$ and $q_{i}=2$.

\subsection{Fermi surface without dipole interaction}
Firstly, we consider the minimal dipole coupling with $p_i=0(i=1,2)$ and discuss the Fermi momentum $k_F$ where the retarded Green functions has a sharp quasi-particle-like peak. Before going on, we would like to point out that the Green function has a symmetry  $G_{22}(\omega,k)=G_{11}(\omega,-k)$ which can
be easily seen from the (\ref{floweqa}) and the boundary condition (\ref{bdycond}). Thus, we mainly focus on studying $G_{22}(\omega,k)$.

\subsubsection{Case I: Only $A_t$ couples with the probe fermions}
\begin{figure}
\centering
\includegraphics[width=.46\textwidth]{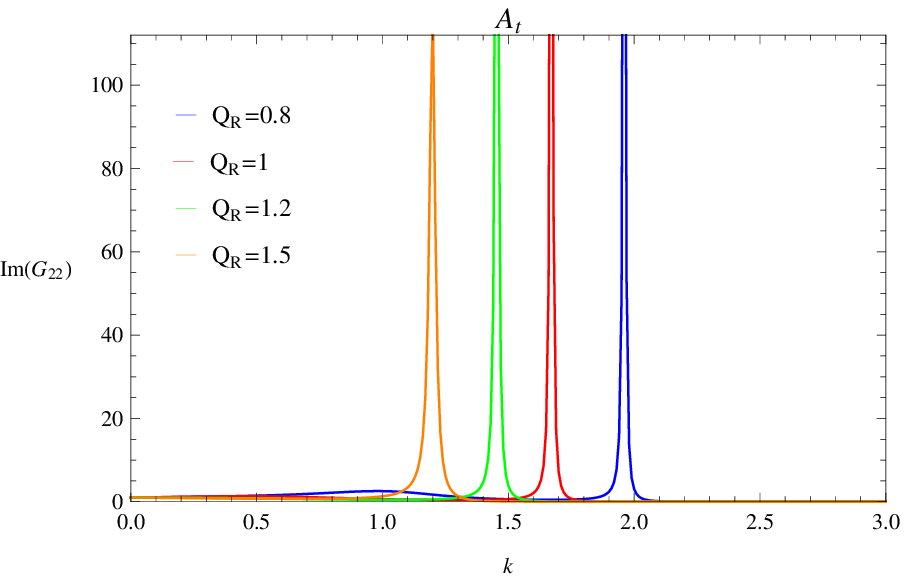}\hspace{0.6cm}
\includegraphics[width=.46\textwidth]{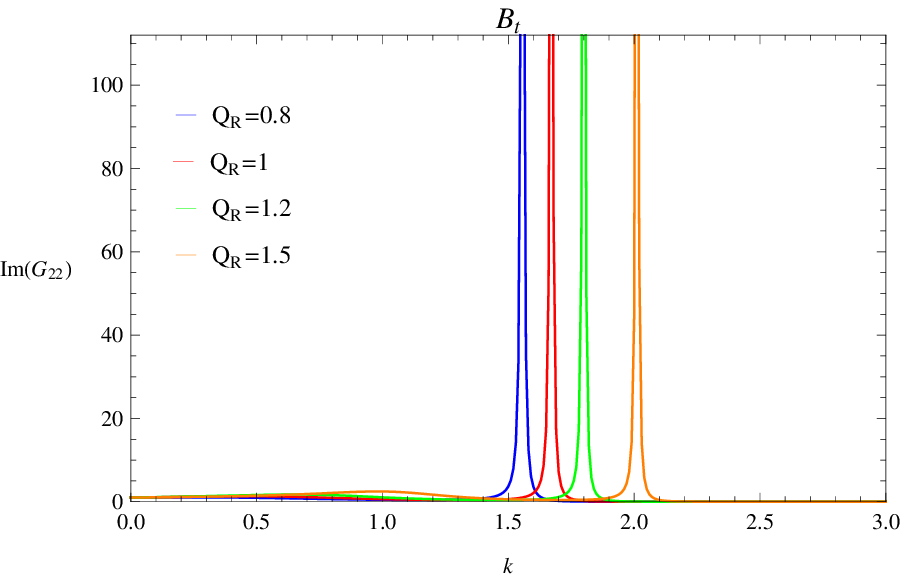}
\caption{The plot of $\rm{Im}G_{22}(k)$  for $Q_{R}=0.8$(blue), $Q_{R}=1$(red), $Q_{R}=1.2$(green) and $Q_{R}=1.5$(orange).}
\label{kfp0}
\end{figure}

Let us  consider the case that only the gauge field $A_t$ interacts with the probe Dirac fermion.  That is to say, we will choose $i=1$ in (\ref{bulkfermionaction}) and (\ref{CovariantDerivative}). For the minimal dipole coupling, we examine the fermi surface carefully and discover that there exists a quasi-particle-like peak for some samples of $Q_R$. Specially, in the left panel of Figure \ref{kfp0}, we shows the behaviors of $\rm {Im}G_{22}(k)$ for some samples of $Q_R$ with $\omega=-10^{-11}$. We see that the Fermi momentum $k_F$ becomes smaller for bigger $Q_R$. The more precise values of $k_F$ are summarized in Table \ref{Atkfz}. When $Q_R<0.5$, $\nu_k$ is imaginary, so the Fermi surface does not exist. This tells us, in the log-oscillatory regime  $Q_R<0.5$ and $p_1=0$, there is no real Fermi surfaces.

In order to estimate the character of dual liquid, we need to check the behavior of $\rm {Im}G_{22}$ near the Fermi momentum $k_{\perp}=k-k_{F}$, named dispersion relation of the Green function $\rm {Im}G_{22}$. The dispersion relation can be expressed as
\begin{equation}\label{dr}
\omega_{\ast}(k_{\perp})\sim k_{\perp}^{\alpha},
\end{equation}
where $\omega_{\ast}(k_{\perp})$ is the distance to the location of Fermi surface and the scaling exponent $\alpha$ can be determined by
\begin{equation}\label{scalexp}
\alpha= \left\{ \begin{array}{ll}
1 & \textrm{~~~when $\nu_{k_{F}}^{I}>\frac{1}{2}$}\\
\frac{1}{2\nu_{k_{F}}^{I}} & \textrm{~~~when $\nu_{k_{F}}^{I}<\frac{1}{2}$}
\end{array} \right.
\end{equation}
with $\nu_{k_{F}}^{I}$ easily obtained from (\ref{nuk}) at $k=k_F$. As we know when $\nu_{k_{F}}^{I}>\frac{1}{2}$ ($\alpha=1$), the dual liquid corresponds to Fermi liquid. When $\nu_{k_{F}}^{I}<\frac{1}{2}$, the dual liquid corresponds to non-Fermi liquid. After determining the Fermi momentum, we analytically calculate the scaling exponent $\alpha$ via (\ref{nuk}) and (\ref{scalexp}). The results in Table \ref{Atkfz} reveal  that the dual holographic system is always non-Fermi liquid as the chemical potential changes. It seems that the change of chemical potential has no influence on the fundamental type of the dual liquid.

\begin{table}
\centering
\footnotesize
\begin{tabular}{|c|c|c|c|c|c|c|c|c|c|}
  \hline
  $Q_{R}$ & 0.5 & 1 & 1.5 & 2 & 2.5 & 3  \\ \hline
  $k_{F}$ & $-$ & 1.6710 & 1.1983 & 0.8674 & 0.6169 & 0.4679  \\ \hline
  $\alpha$& $-$ & 3.37034 & 3.81993 & 5.38319 & 8.4646 & 12.2274  \\
  $(\nu_{k})$& $-$ & (0.148347) & (0.130892) & (0.0928817) & (0.0590695) & (0.0408916)  \\ \hline
\end{tabular}
\caption{$k_F$ and $\alpha$ with different chemical potential at $p_{1}=0$ for $i=1$.}
\label{Atkfz}
\end{table}
\subsubsection{Case II: Only $B_t$ couples with the probe fermions}
Now we turn to study the other case that only the gauge field $B_t$ couples with bulk fermions while $A_t$ is a part of background. The chemical potential $\mu_2$ relative to $B_t$ is not monotonous in the positive region of $Q_R$. The decreasing and increasing interval of $\mu_2$ are the regions of $0<Q_{R}<0.765$ and $Q_{R}>0.765$, respectively. We find that Fermi momentum increases as $Q_R$ increases in the region of $Q_{R}>0.765$ but the rule does not hold in the complementary region. The location of the Fermi surface is shown in the right panel of Figure \ref{kfp0} while the values of $k_F$ and the critical exponents are summarized in Table \ref{Btkfz}. From this numerical result, it seems that different from that in the  previous  case,  and  in this case the Fermi momentum has no consistent rule as the chemical potential changes.  Nonetheless, the similar result to that in the case of nonzero $A_{t}$ is that the type of dual system is also always non-Fermi liquid.

\begin{table}
\centering
\scriptsize
 \begin{tabular}{|c|c|c|c|c|c|c|c|c|c|}
  \hline
  $Q_{R}$ & 0.1 & 0.2 & 0.5 & 0.765 & 1 & 1.5 & 2  \\ \hline
  $k_{F}$ & 1.6870 & 1.4665 & 1.4268 & 1.5377 & 1.6710 & 2.0111 & 2.3975  \\ \hline
  $\alpha$ & 3.7955 & 3.56655 & 3.33173 & 3.31293 & 3.37049 & 3.6601 & 4.15051 \\
  $(\nu_{k})$ & (0.131735) & (0.140191) & (0.150072) & (0.150924) & (0.148347) & (0.136608) & (0.120467)  \\ \hline
 \end{tabular}
\caption{$k_F$ and $\alpha$ with different chemical potential at $p_{2}=0$ for case $i=2$.}
\label{Btkfz}
\end{table}

From the above discussion with the minimal dipole coupling, let us give a short summary: Firstly, the chemical potential has influence on the Fermi momentum when either of the gauge field $A_t$ or $B_t$ interacts with the probe fermion. Secondly, the chemical potential effects the scaling exponent of dispersion relation, but it can not change the type of dual non-Fermi liquid. The system which we study violates the Luttinger's theorem \cite{LHLuttinger}. Because the system has gapless degrees of freedom charged under the $U(1)$ besides the fermions, the charge density is much larger than the volume enclosed by the Fermi surface.

\subsection{Various properties with nonvanishing dipole coupling}
In the above subsection, we do not obtain other phase besides non-Fermi liquid phase. Keeping our motivation in mind, it seems that we shall consider the dipole coupling term. Now we turn on the dipole coupling. We will investigate the influence of the dipole coupling constant $p_{i}$ and chemical potential on the dual system. In order to achieve this goal, we will explore the density of states by integrating $A(\omega,k)$ over $k$. We will use different values of $Q_R$  for discussion in this section since changing $Q_R$ is equivalent to a running chemical potential.
\subsubsection{$A_t$ interacts with the probe fermion}
\begin{figure}
\centering
\includegraphics[width=.39\textwidth]{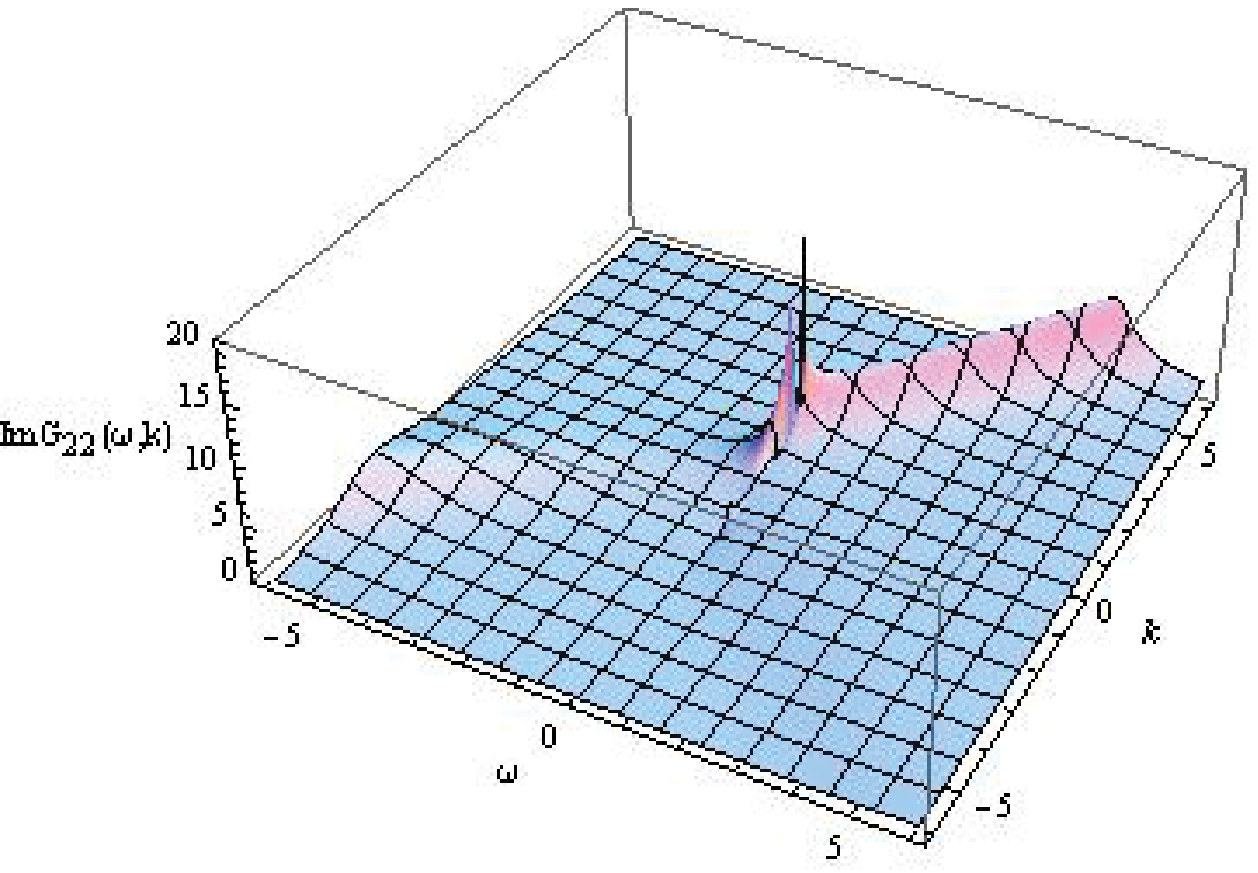}\hspace{1cm}
\includegraphics[width=.39\textwidth]{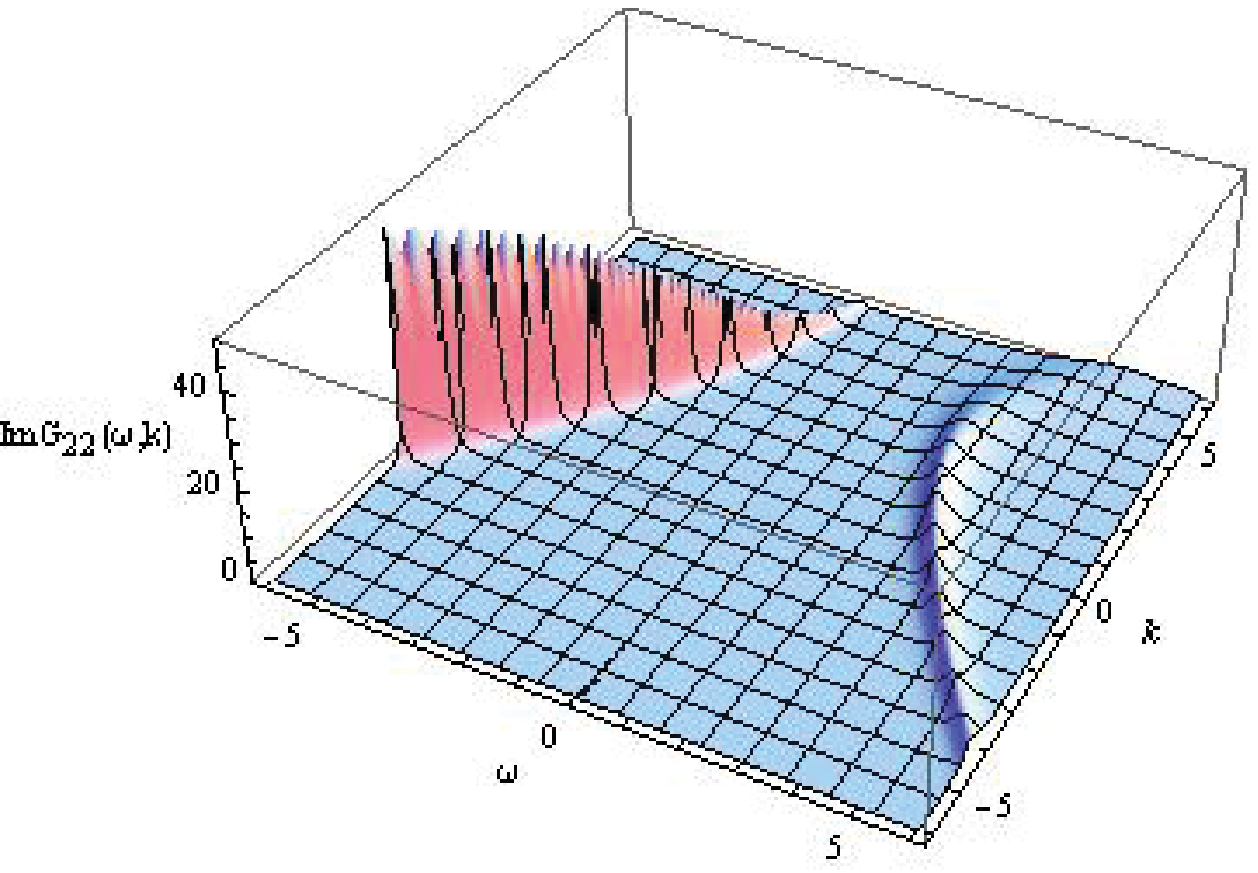}\\
\includegraphics[width=.33\textwidth]{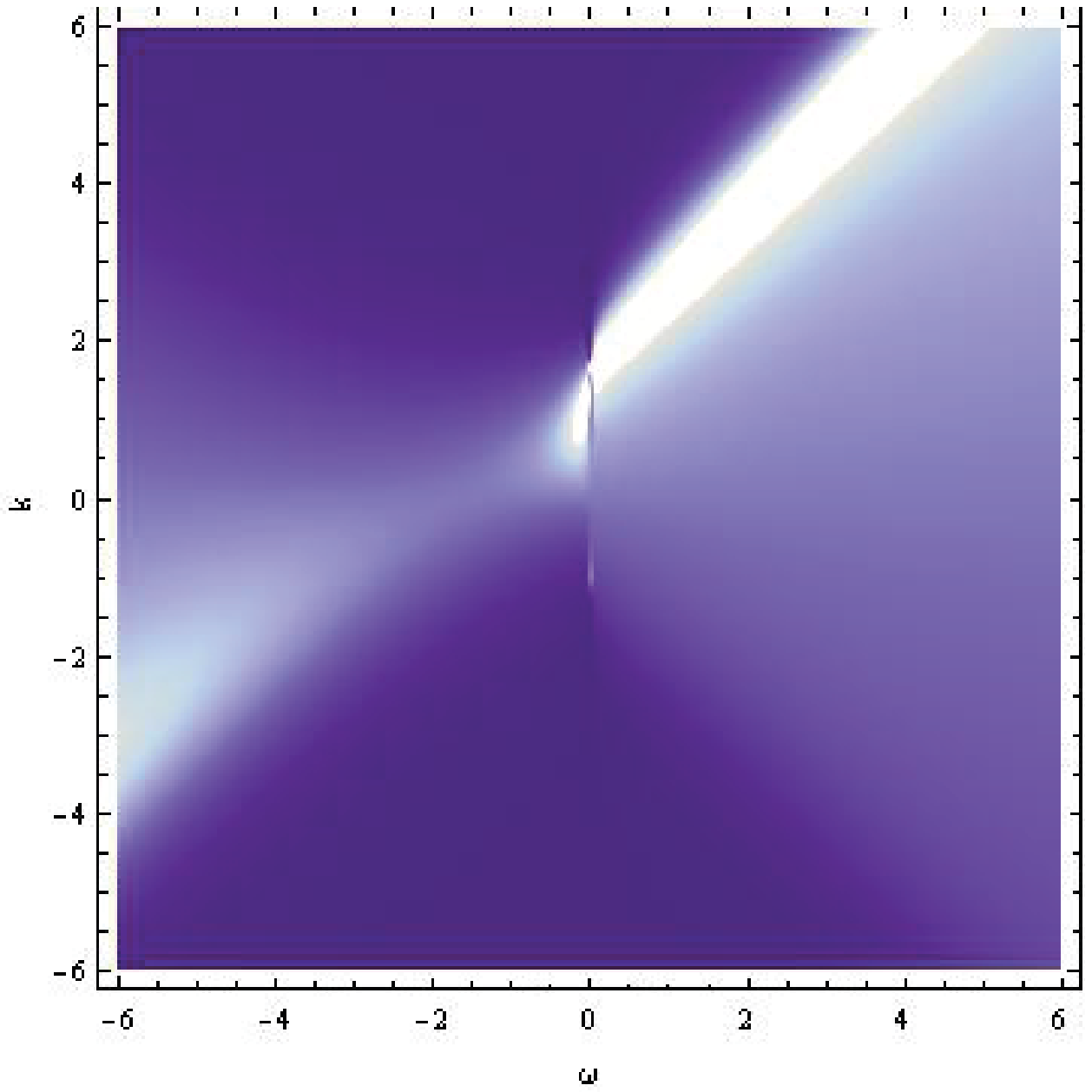}\hspace{1cm}
\includegraphics[width=.33\textwidth]{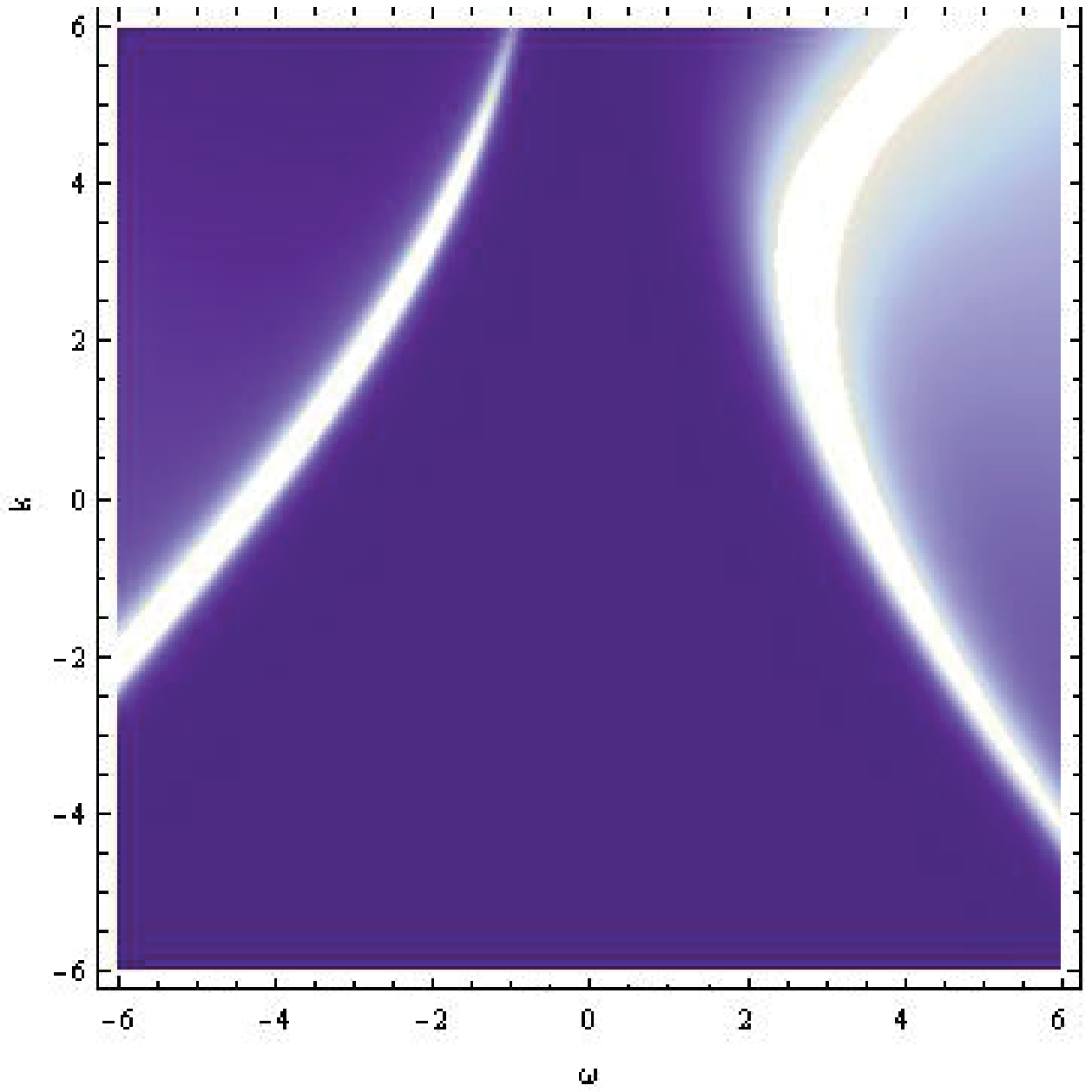}
\caption{The 3D plot of $\rm{Im} G_{22}(\omega,k)$ and its density plot for $p_{1}=0$ (left plane) and $p_{1}=5$ (right plane). We set $q_{1}=2$ and $Q_R=1$.}
\label{gap3d}
\end{figure}

At first, we focus on the case that only $A_t$ couples with the probe fermions. In other word, we will set $i=1$ in (\ref{floweqa}). In figure \ref{gap3d}, the plots show $\rm{Im}G_{22}(\omega,k)$ for $p_{1}=0$(left panel) and $p_{1}=5$(right panel) with $Q_R=1$. The left plots for $p_{1}=0$ show the same result as we analyzed in previous section that the quasi-particle-like peak representing a Fermi surface locates at $k\simeq1.67$ as $\omega\rightarrow0$. The low energy excitations near $k_F$ are non-Fermi liquid type. However, from the right plots for $p_{1}=5$, we find the peak around $\omega\rightarrow0$ disappear and a gap generated.

\begin{figure}
\centering
\includegraphics[width=.5\textwidth]{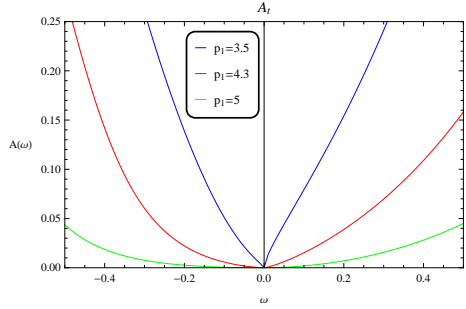}
 \caption{The plot of $A(\omega)$ for $p_{1}=3.5$(blue line), $p_{1}=4.3$(red line) and $p_{1}=5$(green line). The onset of the gap is near $p_{1}=4.3$ with $Q_R=1$.}
 \label{Awqr1p}
\end{figure}

\begin{figure}
\centering
\includegraphics[width=.4\textwidth]{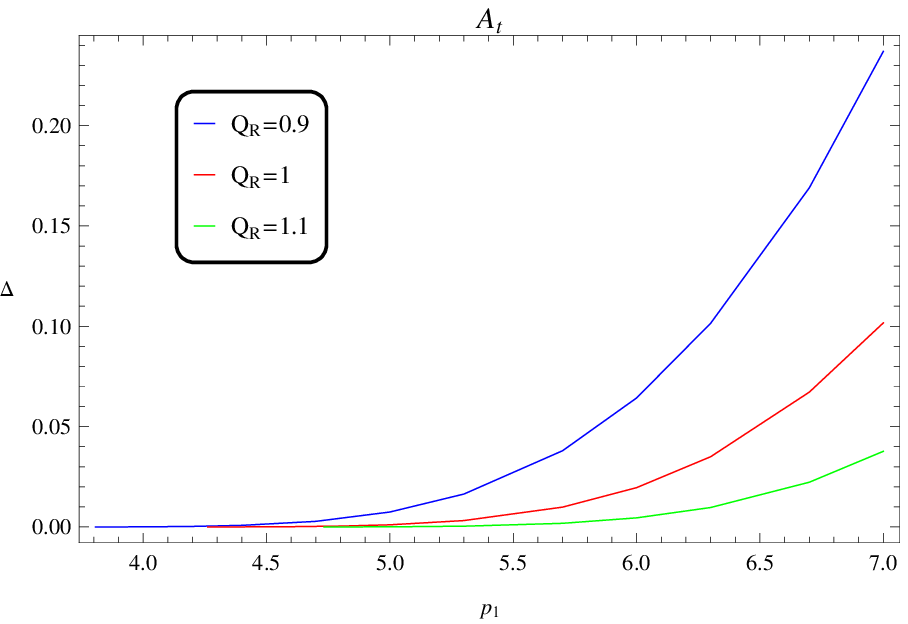}\hspace{0.5cm}
\includegraphics[width=.4\textwidth]{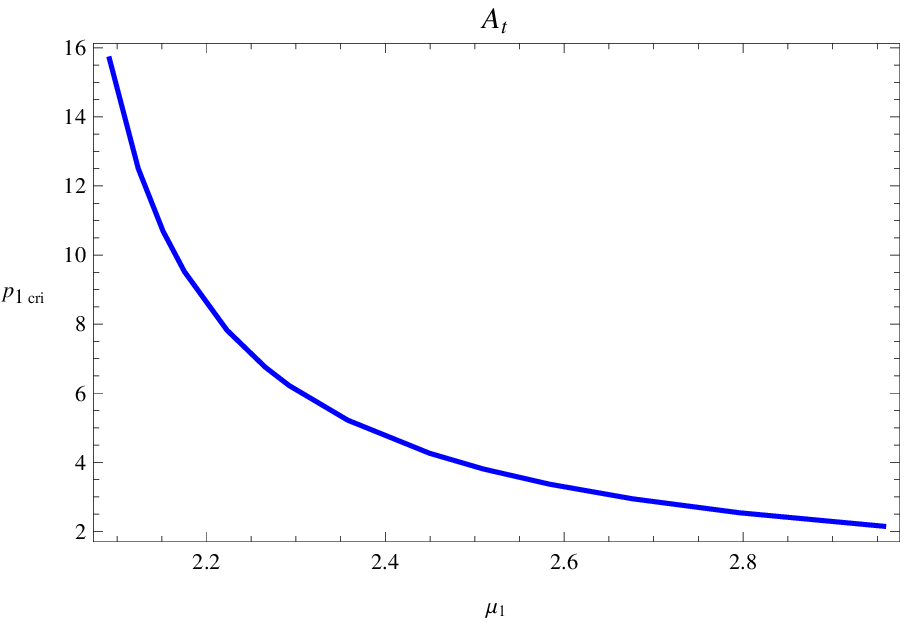}
 \caption{Left: The plot of gap width versus $p_{1}$ for different $Q_R$. Right: $P_{1cri}$ as a function of $\mu_1$.}
 \label{Atgapwidth}
\end{figure}

For small $p_{1}$, the major feature of the system is that the spectral function has a quasi-particle-like peak which locates at $k=k_F$. As $p_{1}$ increases, the peak declines. Once $p_{1}$ reaches a critical value $p_{1cri}$, a gap will emerge. The width of gap will become larger as $p_{1}$ increases further. Also the peaks initiate to transfer from the upper band with $\omega>0$ to the lower band with $\omega<0$. From the state density versus the frequency in figure \ref{Awqr1p}, we can see the onset of gap is near $p_{1cri}=4.3$ with $Q_R=1$\footnote{We integrate $A(\omega)$ over $k$ in sufficiently wide range. When the value of $A(\omega)$ reaches $4.5\times10^{-8}$, we deem the spectral function $A(\omega)\sim0$ accompanying the gap generates at $p_{1cri}$.}. The transfer of the spectral density is also implicit in the figure. Furthermore, the width of the gap enlarging with $p_{1}$ can also be seen in the left plot of figure \ref{Atgapwidth}. In the right plot of figure \ref{Atgapwidth}, we show the effect of the chemical potential on the critical dipole coupling. We see that smaller chemical potential makes the generation of gap at larger dipole coupling. In other words, the relative chemical potential change  the effect of dipole coupling. Other factor that change the dipole coupling effect also discussed in \cite{JPWu1,Wen,Kuang1}.
\begin{figure}
\centering
\includegraphics[width=.39\textwidth]{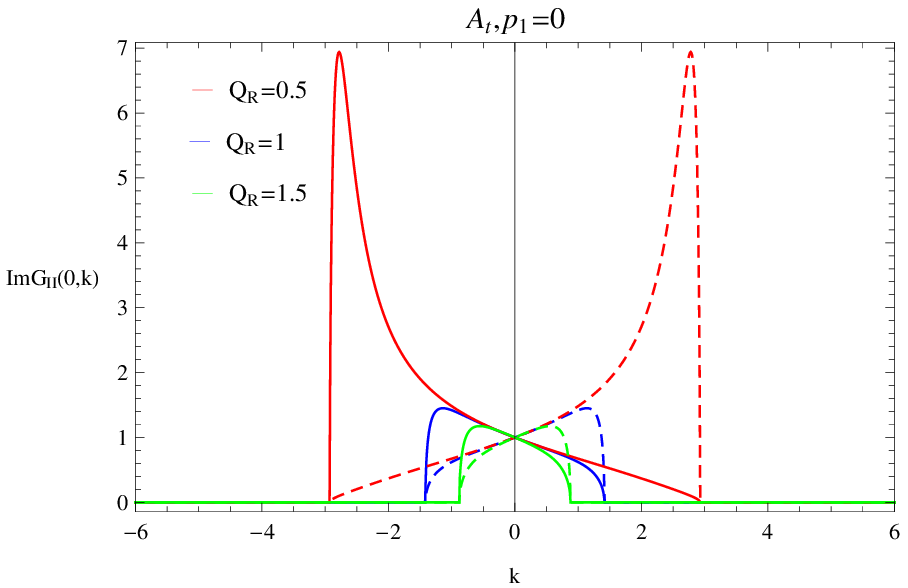}\hspace{1cm}
\includegraphics[width=.39\textwidth]{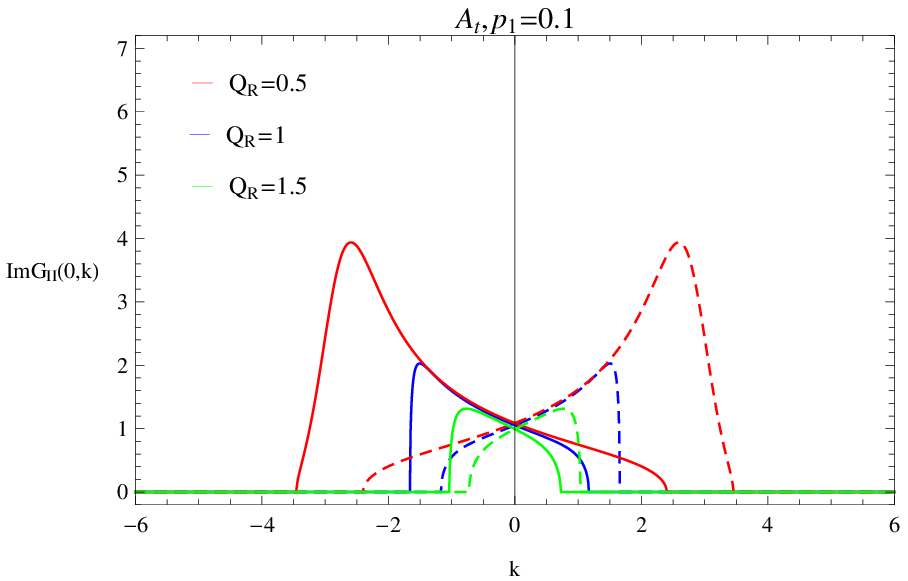}
\caption{$\rm{Im}G_{11}(0,k)$(dashed) and $\rm{Im} G_{22}(0,k)$(solid) for $p_{1}=0$ and $p_{1}=0.1$ with various $Q_R$.}
\label{Atw0kqr}
\end{figure}

In order to see more influence on the holographic fermion system by the running chemical potential, we report our numerical result in the limit when $\omega=0$ with different $Q_R$ in figure \ref{Atw0kqr}. From the figure, it is obvious that the symmetry of $\rm{Im}G_{11}(0,-k)=\rm{Im}G_{22}(0,k)$ is kept for both $p_{1}=0$ and $p_{1}=0.1$. Besides, there is a range of $k$ in which both $\rm{Im}G_{11}(0,k)$ and $\rm{Im}G_{22}(0,k)$ are nonzero. This range of $k$ at fixed $p_{1}$ becomes wider  as $Q_R$ becomes smaller. As pointed out in\cite{f2}, $\rm{Im}G_{II}(\omega,k)$ becomes log-oscillatory when $\omega$ approaches zero in the momentum regime for nonzero $\rm{Im}G_{II}(0,k)$.  The left plot of figure \ref{Atw0kqr} shows that the log-oscillatory regimes coincide at $p_{1}=0$ for all chosen $Q_R$. While this coincidence will shrink for bigger $p_{1}$ as shown in the right plot of figure \ref{Atw0kqr}.

\begin{figure}
\centering
\includegraphics[width=.3\textwidth]{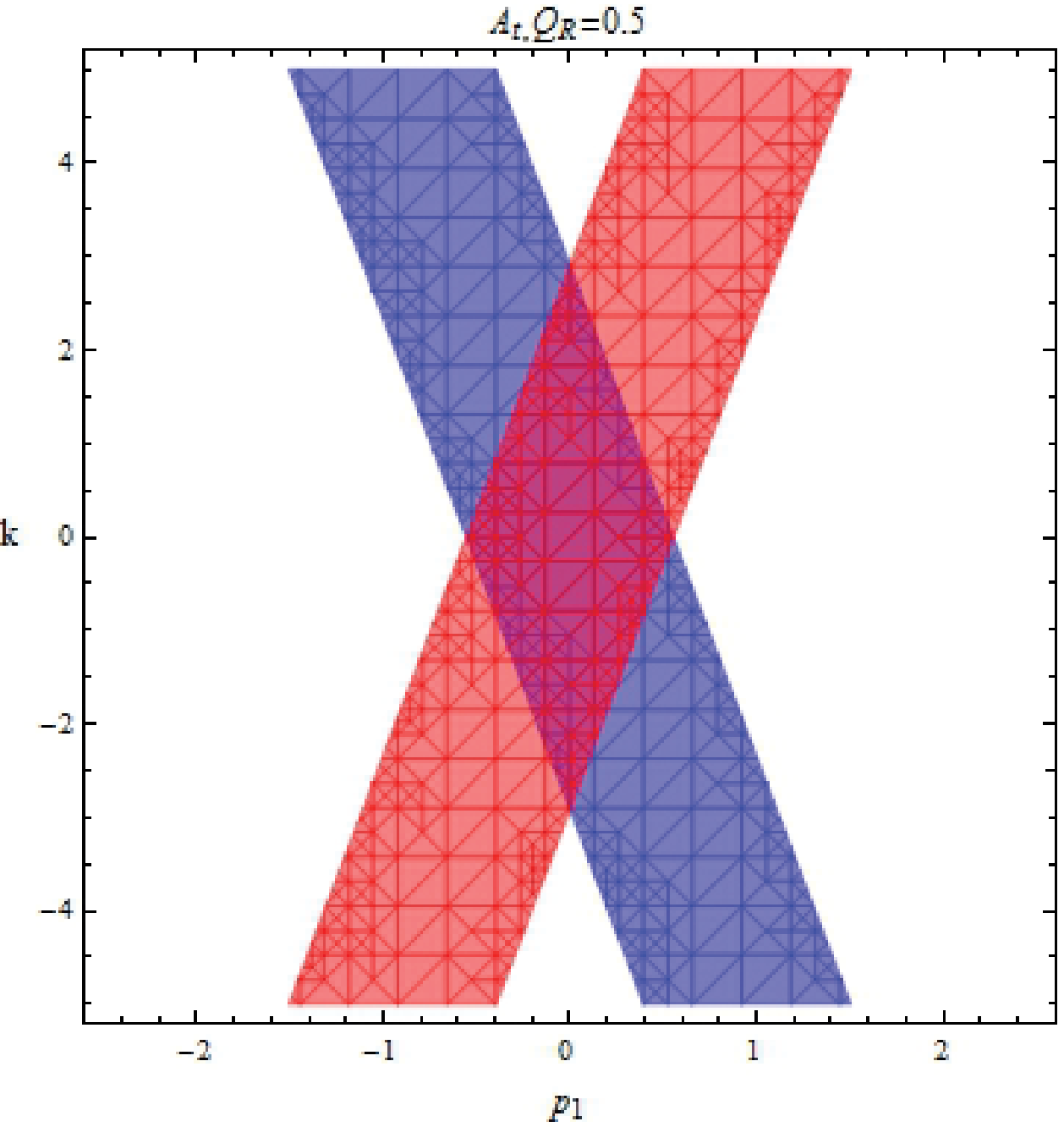}\hspace{0.5cm}
\includegraphics[width=.3\textwidth]{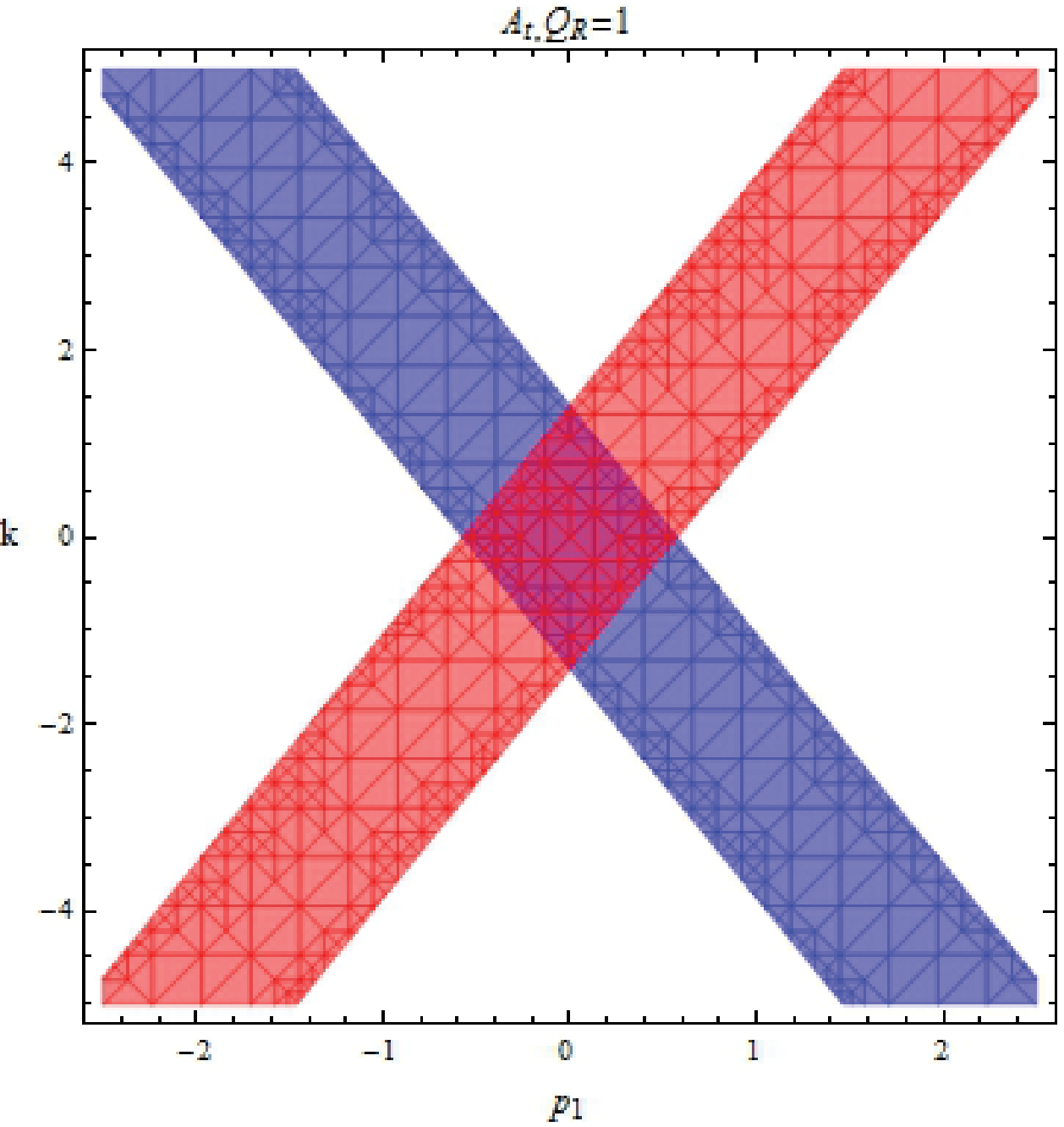}\hspace{0.5cm}
\includegraphics[width=.3\textwidth]{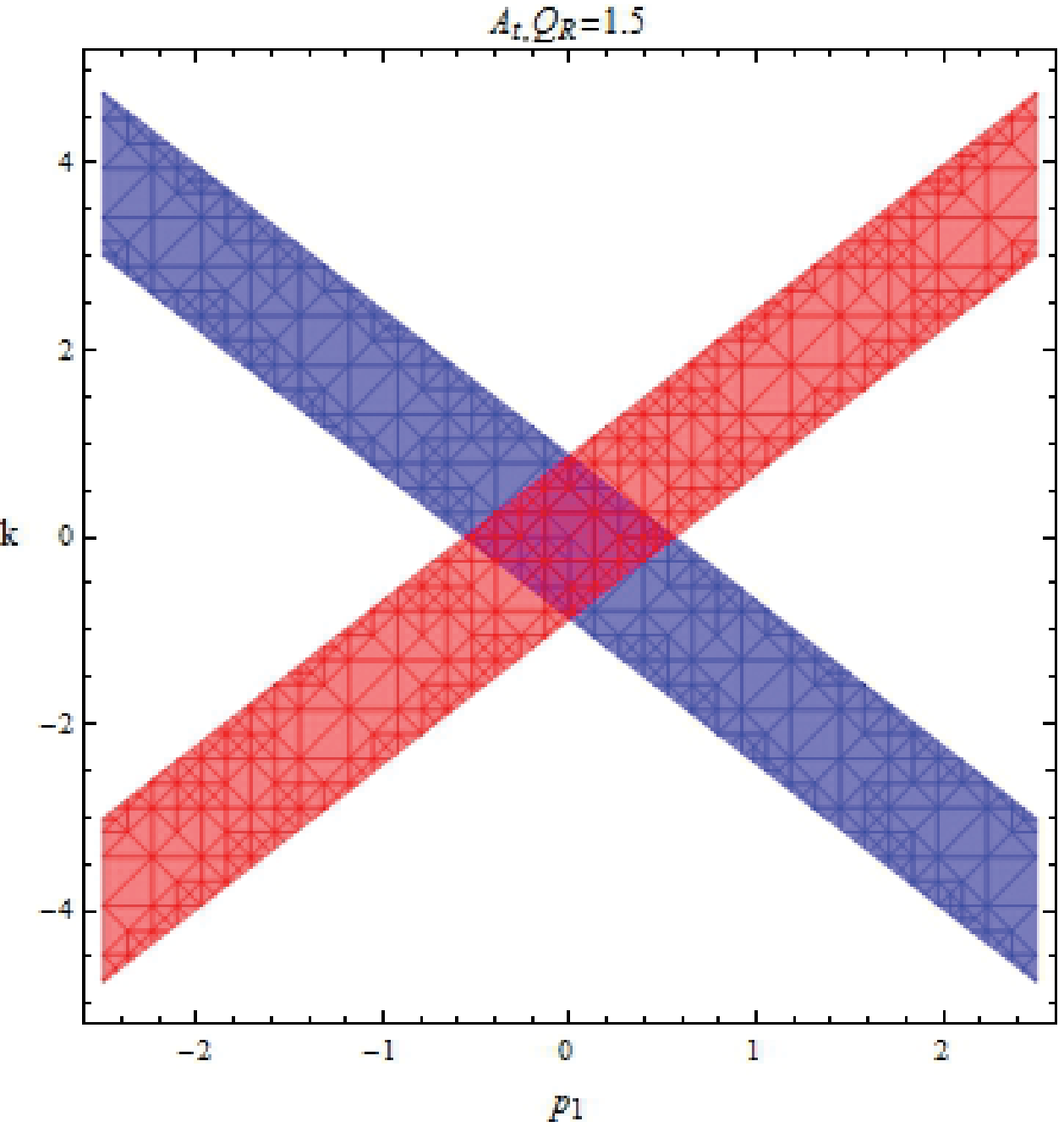}
\caption{The plots of oscillating region for different $Q_R$.}
\label{oscqr}
\end{figure}

Fortunately, we can analytical understand the above properties of the log-oscillatory regimes. By studying the
conformal dimension $\nu_{k}^{I}$ of the dual CFT operator, we find there is a range of momenta $k\in\mathcal{X}_{I}$ in which $\nu_{k}^{I}$ is imaginary.
The momentum regime with imaginary $\nu_{k}^{I}$ corresponds to the log-oscillatory regime\cite{f3} and  Fermi surface do not occur in this regime. In our model,
we have the range of log-oscillatory regime from (\ref{nuk})
\begin{eqnarray}\label{kosc}
k\in\mathcal{X}_{I}=\Big[\frac{(-1)^{I}(\delta_{1i}\beta_{i}k_0 p_{i}+\delta_{2i}\beta_{i}k_0 p_{i})-(\delta_{1i}q_{i} k_{0}\beta_{i}L_{2}+\delta_{2i}q_{i} k_{0}\beta_{i}L_{2})}{L_{2}\tau_0},\nonumber\\
\frac{(-1)^{I}(\delta_{1i}\beta_{i}k_0 p_{i}+\delta_{2i}\beta_{i}k_0 p_{i})+(\delta_{1i}q_{i} k_{0}\beta_{i}L_{2}+\delta_{2i}q_{i} k_{0}\beta_{i}L_{2})}{L_{2}\tau_0} \Big].
\end{eqnarray}
For $k\in\mathcal{X}_{I}$, our boundary condition (\ref{bdycond}) with $\omega=0$ is imaginary while the flow equation is real, so the numerical result in figure \ref{Atw0kqr} that $\rm{Im}G_{II}(0,k)$ is nonzero in this momentum regime is obvious. Also, it is easy to see from (\ref{kosc}) that $\mathcal{X}_{1}=\mathcal{X}_{2}$ for $p_{1}=0$. While for nonzero $p_{1}$, $\mathcal{X}_{1}$ and $\mathcal{X}_{2}$ will be different, both of $\rm{Im}G_{II}(\omega,k)$ are oscillatory only when the values of $k$ belongs to the overlap of $\mathcal{X}_{1}$ and $\mathcal{X}_{2}$. This support
the coincidence and separation of the log-oscillatory regime in figure \ref{Atw0kqr}. The separations of the regimes $\mathcal{X}_{1}$ and $\mathcal{X}_{2}$ for chosen $Q_R$ are plotted in figure \ref{oscqr}, we can see that the smaller chemical potential corresponds to wider log-oscillatory regimes. The analytical results support and explain the numerical result perfectly.
\begin{center}\label{table3}
\begin{table}[ht]
\centering
\scriptsize
\begin{tabular}{|c|c|c|c|c|c|c|c|}\hline
& $p_1=-1$ & $p_1=-0.1$ & $p_1=0$ & $p_1=0.1$  & $p_1=0.2$ & $p_1=0.5$ & $p=5$  \\ \hline
&$k_F=1.02993$ &$k_F=1.68641$& $k_F=1.80341$& $k_F=1.93739$& ~ & ~ & ~\\
$Q_R=0.9$&$\alpha=1$ &$\alpha=2.60958$& $\alpha=3.51576$& $\alpha=6.11777$& NFS & NFS & MI \\
& FL & NFL & NFL & NFL & ~ & ~ & ~ \\ \hline
&$k_F=0.91896$ &$k_F=1.56377$& $k_F=1.67102$& $k_F=1.79031$& $k_F=1.92277$ & ~& ~\\
$Q_R=1$&$\alpha=1$ &$\alpha=2.66054$& $\alpha=3.37034$& $\alpha=4.81534$& $\alpha=13.0162$& NFS & MI \\
& FL & NFL & NFL & NFL & NFL & ~ & ~ \\ \hline
&$k_F=0.81924$ &$k_F=1.45591$& $k_F=1.55604$& $k_F=1.66506$ & $k_F=1.78357$& ~ & ~\\
$Q_R=1.1$ &$\alpha=1.09611$& $\alpha=2.74919$& $\alpha=3.34608$& $\alpha=4.37043$ & $\alpha=6.90966$ & NFS & MI \\
& NFL & NFL & NFL & NFL & NFL & ~ & ~\\ \hline
\end{tabular}
\caption{ The Fermi momentum $k_F$ and the exponent $\alpha$ of dispersion relation with different
$p$ and various $Q_R$ for $A_t$. NFS means the system doesn't present Fermi surface. FL, NFL and MI denote Fermi liquid, non-Fermi liquid and Mott insulator, respectively.}
\end{table}
\end{center}

We summarize  the phase structure of the holographic fermions coupled with $A_t$ field in table \ref{table3}. When we change the value of $p_1$ from small positive to negative, the excitation near the Fermi surface can change from non-Fermi Liquid to Fermi Liquid. Note that ``NFS" means $\nu_k$ is imaginary where Fermi surface do not occur. We also find that when $p_1=-1$, decreasing chemical potential can make the dual Fermi liquid system change to non-Fermi liquid.
\subsubsection{$B_t$ interacts with the probe fermion}
\begin{figure}
\centering
\includegraphics[width=.5\textwidth]{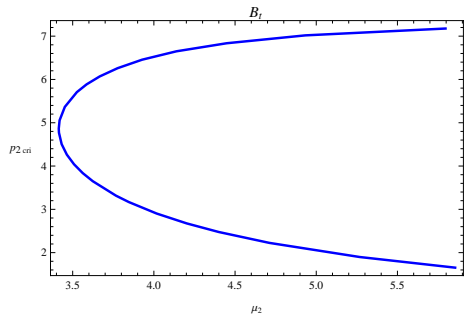}
 \caption{$P_{2cri}$ as a function of $\mu_2$ for the $B_t$ case.}
 \label{Btmupcri}
\end{figure}

\begin{figure}
\centering
\includegraphics[width=.39\textwidth]{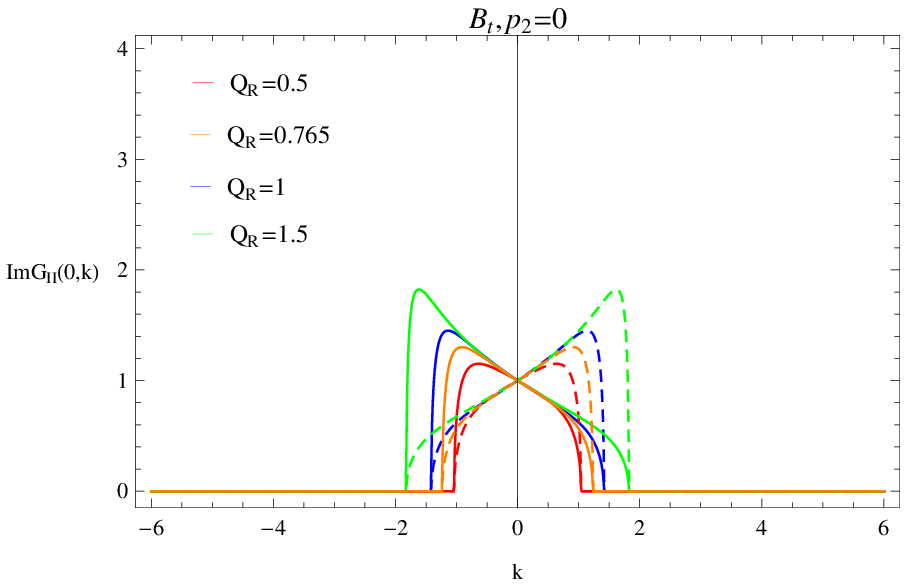}\hspace{1cm}
\includegraphics[width=.39\textwidth]{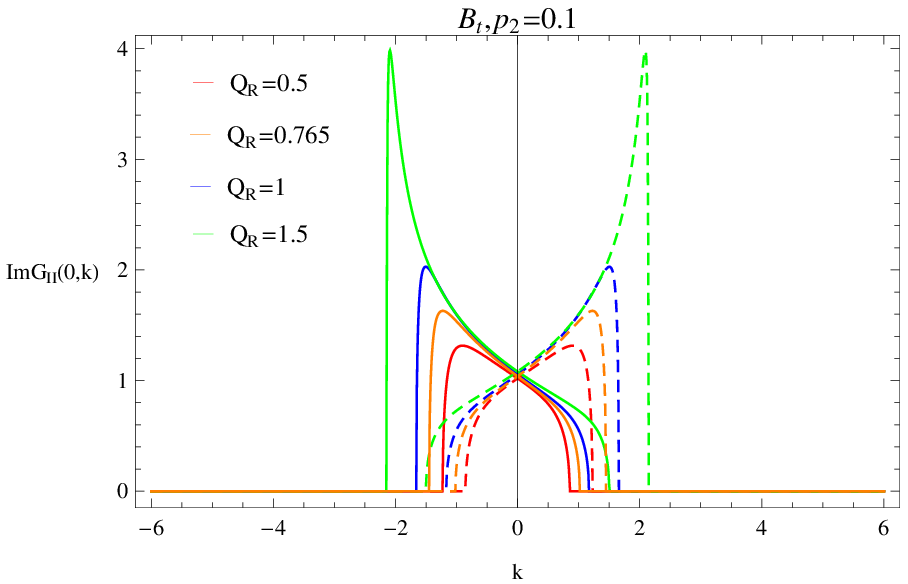}
\caption{$\rm{Im}G_{11}(0,k)$(dashed) and $\rm{Im}G_{22}(0,k)$(solid) for $p_{2}=0$ and $p_{2}=0.1$ with various $Q_R$ for $B_t$ case.}
\label{Btw0kqr}
\end{figure}

\begin{figure}
\centering
\includegraphics[width=.23\textwidth]{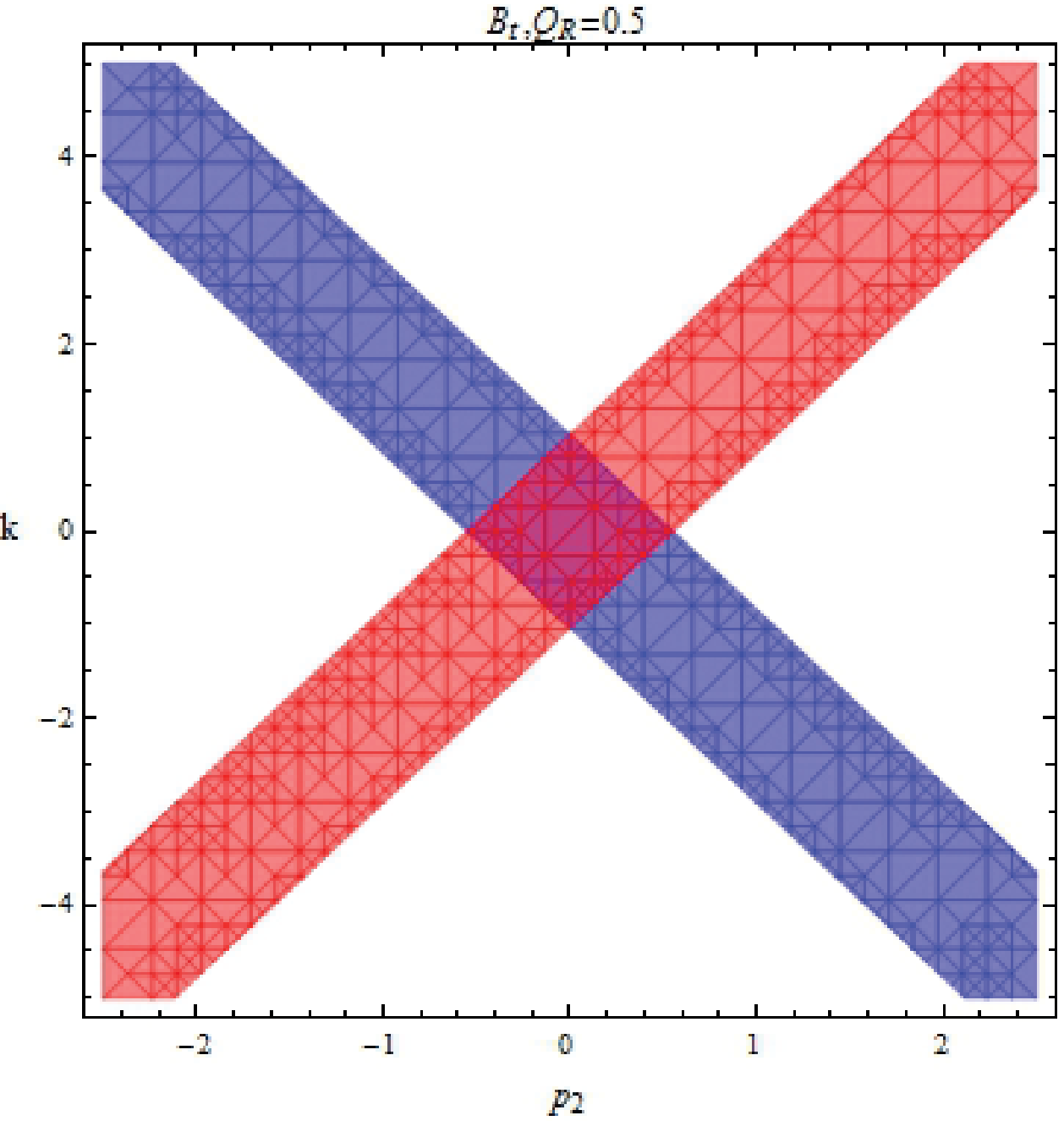}\hspace{0.1cm}
\includegraphics[width=.23\textwidth]{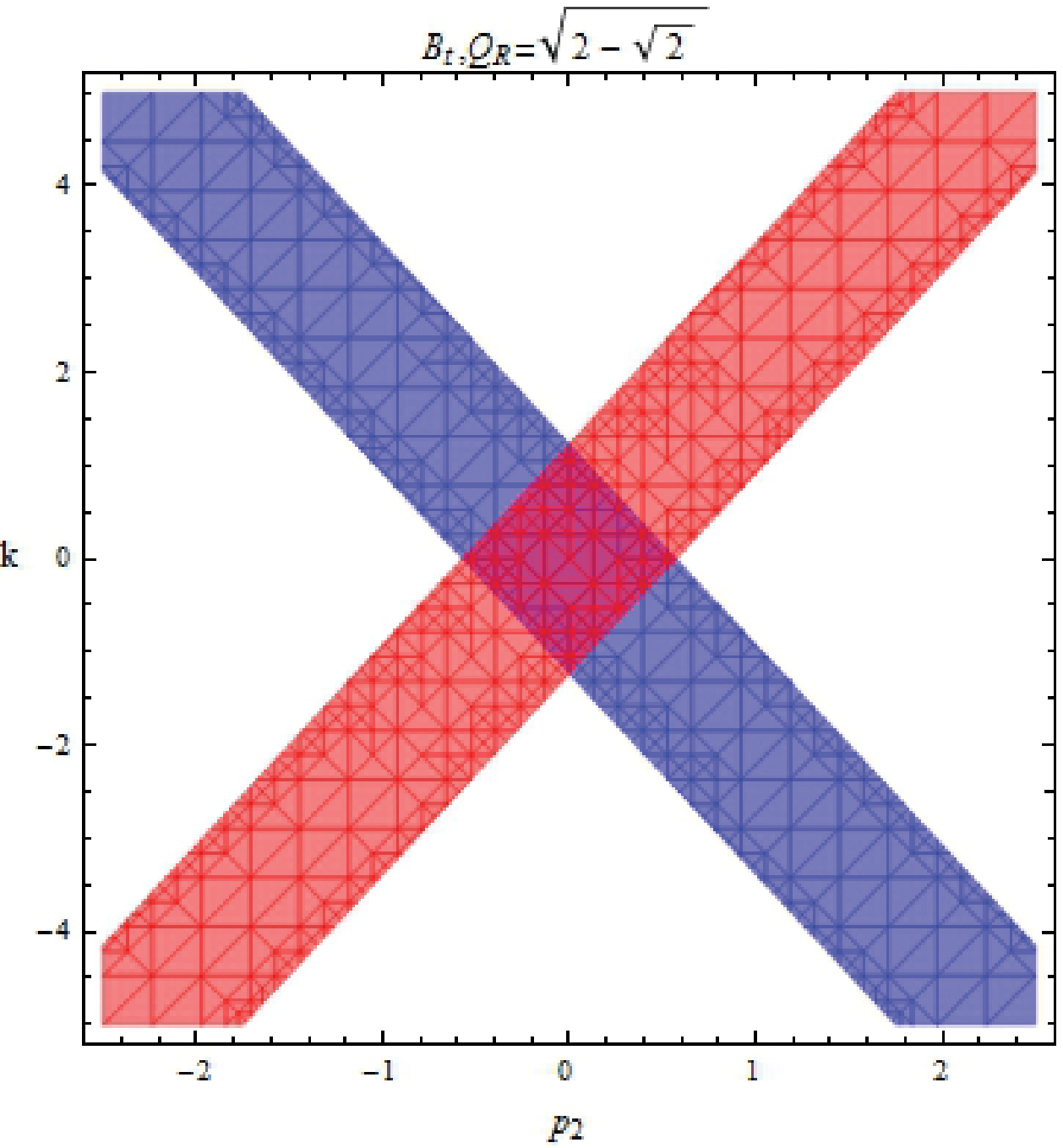}\hspace{0.1cm}
\includegraphics[width=.23\textwidth]{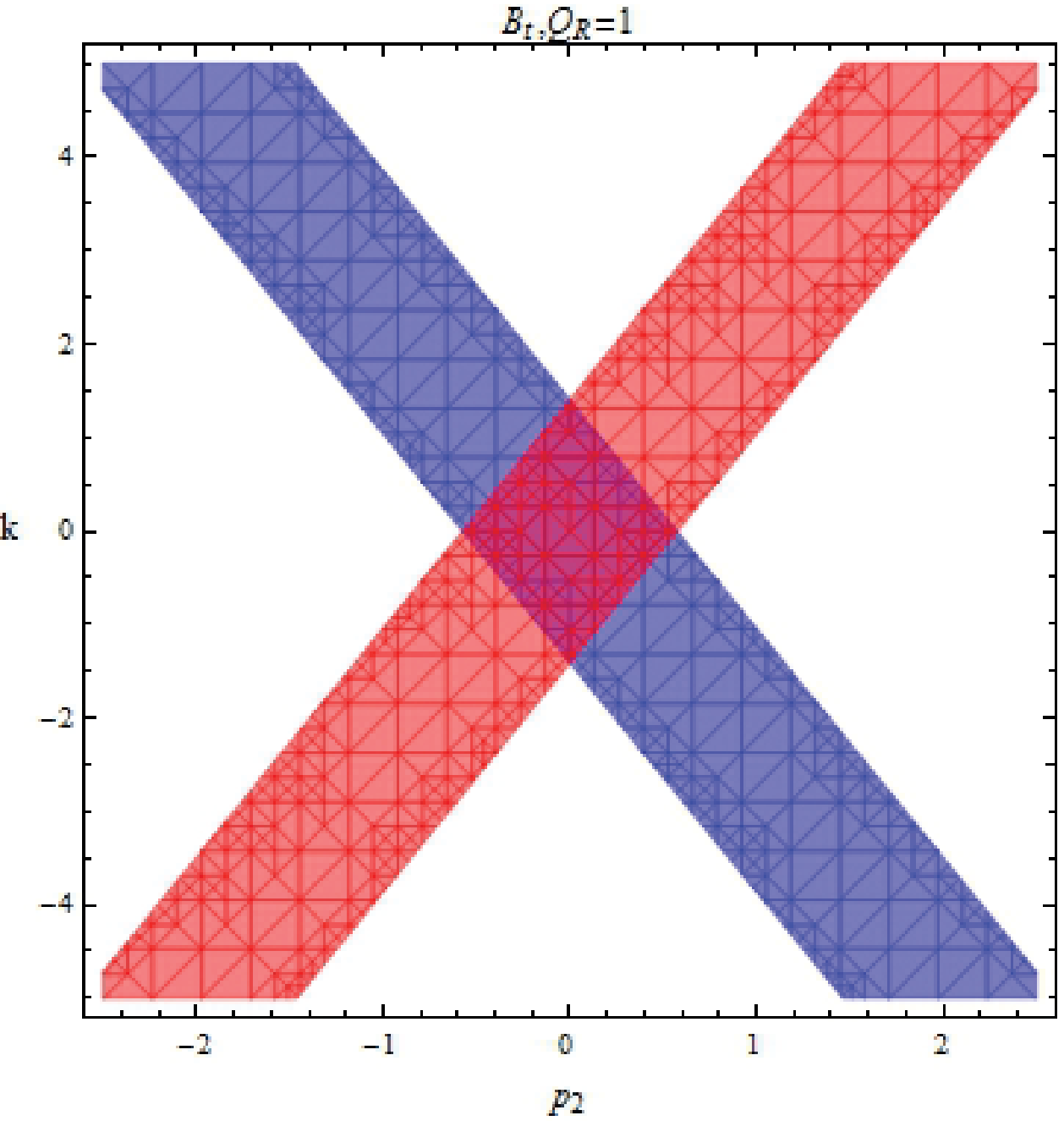}\hspace{0.1cm}
\includegraphics[width=.23\textwidth]{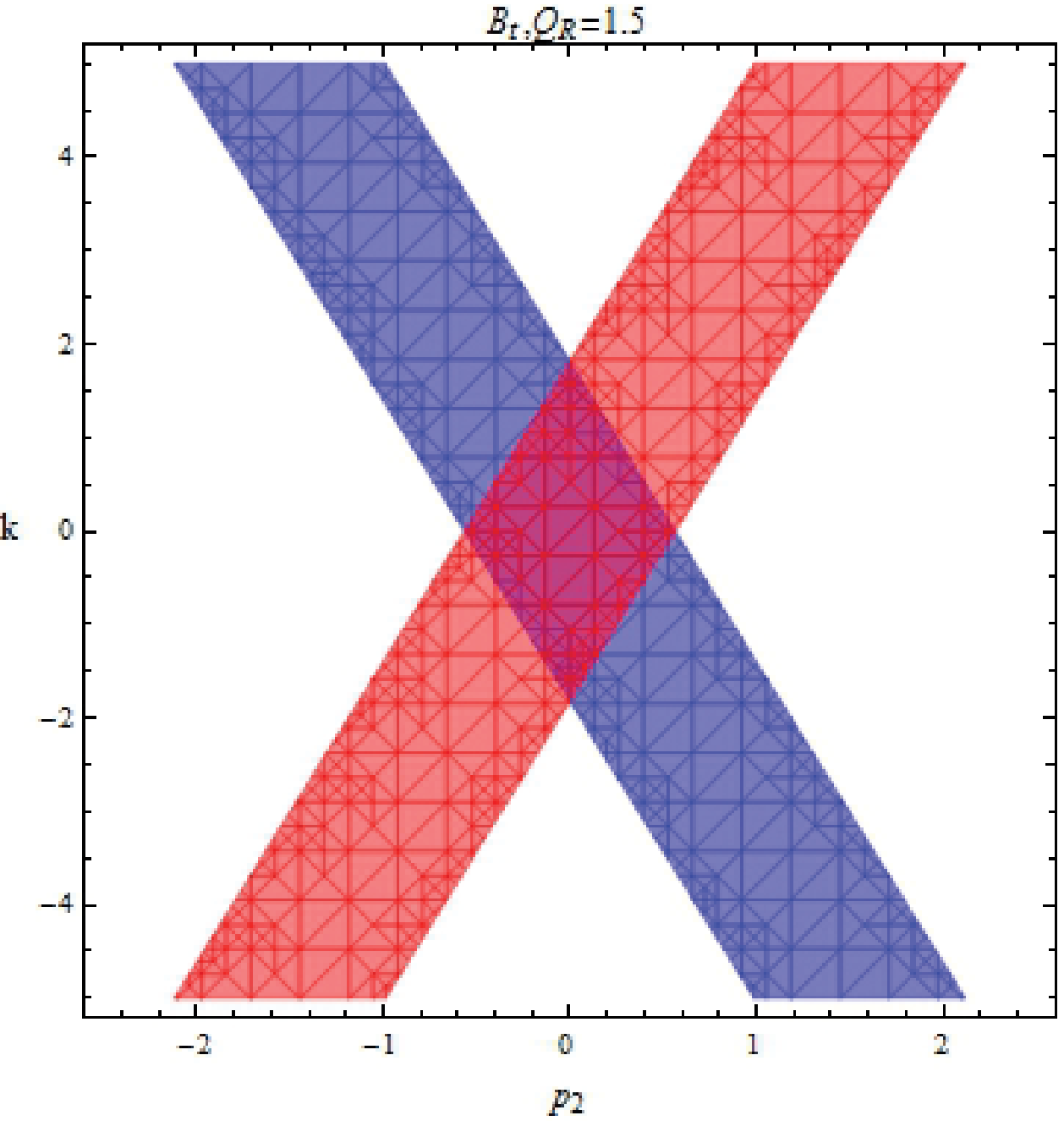}
\caption{The plots of oscillating region with different $Q_R$ for $B_t$ case. }
\label{Btoscqr}
\end{figure}

We also do some parallel work in the case that only  $B_t$ couples with bulk fermions (i.e. i=2). The results are shown in figure \ref{Btmupcri}-\ref{Btoscqr}.  To no one's surprise, there also exists Fermi surface.  In figure \ref{Btmupcri}, $p_{cri}$ of the onset of the gap will change with the change of the chemical potential.
 But the features are quite different from that in the case of $\mu_1$, because $\mu_2$ is a quadratic function of the
 chargeless ratio $Q_R$.  So $\mu_2$ is also a quadratic function of $p_{cri}$.

The properties of the log-oscillatory regimes is also worth studying. Figure \ref{Btw0kqr} is the numerical result given by using the $\omega=0$ boundary condition (\ref{bdycond}). It shows the log-oscillatory regimes is narrower when $Q_R$ increase. Figure \ref{Btoscqr} is the corresponding analytical result, which also agrees well with the numerical results on the width of the log-oscillatory regimes. The separation of the regimes $\mathcal{X}_{1}$ and $\mathcal{X}_{2}$ is smaller as $Q_R$ enlarges.
\begin{center}\label{table4}
\begin{table}[ht]
\tiny
\begin{tabular}{|c|c|c|c|c|c|c|c|}\hline
& $p_2=-1$ & $p_2=-0.1$ & $p_2=0$ & $p_2=0.1$  & $p_2=0.2$ & $p_2=0.5$ & $p=5$ \\ \hline
&$k_F=0.86751$ &$k_F=1.50811$& $k_F=1.61142$& $k_F=1.72495$& $k_F=1.84952$ & ~ & ~\\
$Q_R=0.9$&$\alpha=1.04006$ &$\alpha=2.69864$& $\alpha=3.33878$& $\alpha=4.51229$& $\alpha=8.16924$ & NFS & MI \\
& NFL & NFL & NFL & NFL & NFL & ~ & ~\\ \hline
&$k_F=0.91896$ &$k_F=1.56377$& $k_F=1.67102$& $k_F=1.79031$& $k_F=1.92277$ & ~ & ~\\
$Q_R=1$&$\alpha=1$ &$\alpha=2.66054$& $\alpha=3.37034$& $\alpha=4.81534$& $\alpha=13.0162$& NFS & MI  \\
& FL & NFL & NFL & NFL & NFL & ~ & ~\\ \hline
&$k_F=0.97232$ &$k_F=1.62246$& $k_F=1.73398$& $k_F=1.85959$ & ~ & ~&~\\
$Q_R=1.1$ &$\alpha=1$& $\alpha=2.62679$& $\alpha=3.41129$& $\alpha=5.21841$ & NFS & NFS & MI \\
& FL & NFL & NFL & NFL & ~ & ~ &~\\ \hline
\end{tabular}
\caption{ The Fermi momentum $k_F$ and the exponent $\alpha$ of dispersion relation with different
$p$ and various $Q_R$ for $B_t$.}
\end{table}
\end{center}

The phase structure of holographic fermions coupled with only $B_t$ field is shown in table \ref{table4}. We also find the similar phenomena with the $A_t$ case that bigger charge ratio $Q_R$ can help the type of the dual system change from Fermi liquid to non-Fermi liquid.

\section{Conclusions and Discussions}
Inspired by the phase diagram of the high $T_c$ cuprates, we have shown the influence of the running chemical potential on the dual fermion system in the R-charged  geometry. In the minimal dipole coupling, the running chemical potential directly proportional to the Fermi momentum both for two cases (i.e. gauge fields $A_t$ or $B_t$ couples with fermions, respectively). No matter what changes on the  chemical potential, the properties of the dual system are the non-Fermi liquid type. As to the Dirac operator coupled with only the $A_t$ field case, the bigger chemical potential makes the generation of the gap easier. When only $B_t$ field interacts with the bulk fermions, we found that $\mu_2$ is a quadratic function of the critical dipole coupling constant $p_{2 cri}$. We noted that  $\mu_1$ corresponds to one black hole charge  case and $\mu_2$ corresponds to two equal charges, which probably play a role in the diagrams of the $\mu_i$-$p_{i~cri}$ relation. But the specific reason need us to do further research. We obtained the phase structure  for the holographic fermions for both the  $A_t$ and $B_t$ cases with dipole coupling.  We found that when we turn on dipole coupling, the running chemical potential can influence the type of the dual liquid. In summary, the presence of the running chemical potential and the dipole coupling enrich the physical picture of the holographic fermions.
\section*{Acknowledgements}
We would like to thank Bin Wang for encouragement.
The work was partly supported by NSFC (No. 11075036, No.11375110 and No. 11005072). XHG was
also partly supported by Shanghai Rising-Star Program (No.10QA1402300).

\end{document}